\documentclass[12pt]{article}

\usepackage{amsmath,amssymb}
\renewcommand{\theequation}{\thesection.\arabic{equation}}
\def\be{\begin{equation}}
\def\ee{\end{equation}}
\def\bea{\begin{eqnarray}}
\def\eea{\end{eqnarray}}
\def\kslash{\slash \hspace{-0.2cm} k}
\begin{document}
\thispagestyle{empty}

\begin{flushright}
KEK-TH-1006 \\
OIQP-05-01
\end{flushright}

\vskip 5mm
\begin{center}
{\Large\bf   Fermionic Backgrounds and Condensation of Supergravity Fields
in IIB Matrix Model\\}
\vskip 10mm

{\large
 Satoshi Iso$^{*}$\footnote{\tt satoshi.iso@kek.jp},
 Fumihiko Sugino$^\dagger$\footnote{\tt fumihiko\_sugino@pref.okayama.jp},
 Hidenori Terachi$^{*}$\footnote{\tt terachi@post.kek.jp}
 and Hiroshi Umetsu$^\dagger$\footnote{\tt hiroshi\_umetsu@pref.okayama.jp}
} 
\vskip 5mm
$^*${\it Institute of Particle and Nuclear Studies \\
High Energy Accelerator Research Organization (KEK) \\
Oho 1-1, Tsukuba, Ibaraki 305-0801, Japan \\
and \\
Department of Particle and Nuclear Physics, \\
the Graduate University  for Advanced Studies (Sokendai)\\
Oho 1-1, Tsukuba, Ibaraki 305-0801, Japan} 

\vspace{5mm}

$^\dagger${\it Okayama Institute for Quantum Physics \\
 1-9-1 Kyoyama, Okayama 700-0015, Japan} \\
\end{center}

\vskip 10mm
\centerline{{\bf{Abstract}}}
\vskip 3mm

In a previous paper \cite{ITU} we constructed wave functions of a D-instanton 
and  vertex operators in type IIB matrix model by expanding supersymmetric
Wilson line operators. 
They describe couplings of a D-instanton and type IIB matrix model to the
massless 
closed string fields respectively and form a  multiplet of 
$D=10$ ${\cal N}=2$  supersymmetries.
In this paper we consider fermionic backgrounds and condensation of 
supergravity fields in IIB matrix model by using these wave functions.
We start from the  type IIB matrix model 
in a flat background 
whose matrix size is $(N+1) \times (N+1)$, or equivalently
the effective action for $(N+1)$ D-instantons. 
We then  calculate an effective action for 
$N$ D-instantons by integrating out one D-instanton
(which we call a mean-field D-instanton)  with an appropriate wave function
and show that various terms can be induced corresponding to the choice of
the wave functions. 
In particular, a Chern-Simons-like term is induced when the mean-field
D-instanton has  
a wave function of the antisymmetric tensor field.  A fuzzy sphere becomes 
a classical solution to the equation of motion for the effective action.

We also give an interpretation of the above wave functions in the
superstring theory side as  overlaps of the D-instanton boundary state with
the closed string massless states in the Green-Schwarz formalism.

\newpage

\setcounter{page}{1}
\setcounter{footnote}{0}
\setcounter{equation}{0}
\parskip 1mm

\section{Introduction}

Type IIB (IKKT) matrix model has been proposed as a nonperturbative 
formulation of superstring theory of type IIB~\cite{IKKT}. 
As an evidence for the nonperturbative formulation,
the Schwinger-Dyson equation of Wilson lines is shown to describe
the string field equation of motion of type IIB superstring 
in the light-cone gauge under some plausible assumptions about the continuum
limit~\cite{FKKT}.
Although there are still many issues to be resolved, the model has 
an advantage to other formulations of superstrings that we can discuss 
dynamics of space-time more directly
\cite{AIKKT, Nishimura-Sugino, Kawai-1, Kawai-2}.
The action of the model is given by
\be
S_{\mathrm{IKKT}} = -\frac{1}{4}\mathrm{tr}~[A_\mu,A_\nu]^2 
- \frac{1}{2} \mathrm{tr}~\bar{\psi}\Gamma^\mu [A_\mu,\psi],
\label{IKKT}
\ee
where $A^{\mu}$ ($\mu =0, \cdots,9$) and ten-dimensional Majorana-Weyl 
fermion $\psi$ are $N \times N$ bosonic and fermionic hermitian matrices. 
The action was originally derived from the Schild action for the 
type IIB superstring by regularizing the world sheet coordinates by matrices.
It is interesting that 
the same action describes the effective action for $N$ D-instantons
\cite{Witten}.
This suggests a possibility that D-instantons  can be considered 
as fundamental objects to generate the space-time itself as well as 
the dynamical fields (or strings) on the space-time.
The bosonic matrices represent noncommutative coordinates of D-instantons and 
the distribution of eigenvalues of $A_\mu$ is interpreted to form space-time.
The fermionic coordinates $\psi$ are collective coordinates associated with
broken supersymmetries of D-branes but in the matrix model interpretation
they describe internal structures of our space-time.

If we take the above interpretation that the space-time is constructed 
by distribution of D-instantons, how can we interpret the $SO(9,1)$ 
rotational symmetry of the matrix model action?
This symmetry can be interpreted in the sense of mean-field.
Namely we can consider that the system of $N$ D-instantons is embedded in
larger size 
$(N+M)\times (N+M)$ matrices as 
\be
\left(
\begin{array}{cc}
N D(-1) & \\
& M D(-1) \mbox{ as background for } N D(-1) 
\end{array} \right),
\label{NM}
\ee
and consider the action (\ref{IKKT}) as an effective action 
in the background where the rest, $M$ eigenvalues, distribute uniformly 
in 10 dimensions.
If the $M$ eigenvalues distribute inhomogeneously,
we may expect that the effective action for $N$ D-instantons is modified
so that they live in a curved space-time.
This is analogous to a thermodynamic system. In a canonical ensemble,
a subsystem in a heat bath is characterized by several thermodynamic 
quantities like temperature and pressure. Similarly a subsystem 
of $N$ D-instantons in a ``matrix bath'' can be considered to be
characterized by several ``thermodynamic quantities'' in a certain large $N$
limit. 

Since the matrix model has the ${\cal N}=2$ type IIB supersymmetry
\begin{eqnarray}
\left\{
\begin{array}{rl}
\delta A_\mu =& i \bar{\epsilon}_1 \Gamma_\mu \psi, \\
\delta \psi =& \displaystyle{
 -\frac{i}{2} [A_\mu , A_\nu] \Gamma^{\mu\nu} \epsilon_1
 }
 + \epsilon_2 1_{N},
\end{array}
\right. 
\end{eqnarray}
we can expect that the configuration of the $M$ background D-instantons 
describes condensation of massless fields of the type IIB supergravity 
and the thermodynamic quantities of the matrix bath are characterized by the
values of the condensation.

In the following, we consider the simplest case that the background 
is represented by a wave function of one D-instanton 
(namely $M=1$ with an appropriate wave function introduced).
This simplification can be considered as a mean-field approximation 
that the configuration of $M$ D-instantons is represented by a  
mean-field wave function described by a single D-instanton.
We call this extra D-instanton a {\bf mean-field D-instanton}.
This kind of idea  was first discussed by Yoneya in \cite{Yoneya}. 
In the previous paper \cite{ITU}, we  constructed 
a set of wave functions for the mean-field D-instanton. 
This wave functions have a stringy interpretation,
namely, as we see in this paper, they can be
interpreted as overlaps of D-instanton boundary states with closed
string massless states.

In this paper, we calculate the effect of the mean-field D-instanton on the 
$N$ D-instantons. We first start from the IIB matrix model
with a size $(N+1)\times(N+1)$, or 
equivalently a system of $(N+1)$ D-instantons,  and integrate 
the mean-field D-instanton with an appropriate wave function.
This corresponds to  condensation of supergravity fields and
the effective action for the $N$ D-instantons is modified.
We particularly consider two types of wave functions, namely 
those describing 
an antisymmetric tensor field or a graviton field.
In the former case, a Chern-Simons like term is induced in the leading order
of the perturbation. (This term vanishes if we assume that the $N$
D-instantons satisfy the equation of motion for the original IIB action.)
With this term, a fuzzy sphere becomes a solution 
to the equation of motion. This phenomenon is similar to the Myers
effect~\cite{Myers}. 
In both cases for the antisymmetric tensor field and the graviton, 
if we assume that the configuration satisfies the classical
equation of motion, the modification of the effective action is given by 
a vertex operator for each supergravity field.

The content of the paper is as follows.
In the next section, we review the results of the previous paper \cite{ITU}
on the wave functions of a D-instanton and the vertex operators
for closed string massless states in IIB matrix model.
In section 3,   we give a stringy interpretation of the D-instanton
wave functions as overlaps of D-instanton boundary states with
massless states of the closed string.
In section 4, we calculate the one-loop effective action in 
general fermionic backgrounds. In section 5, we apply this
calculation to a system of $(N+1)$ D-instantons 
and integrate the mean-field D-instanton to
obtain an effective action for the rest $N$ D-instantons.
We particularly consider the wave functions of the antisymmetric tensor
field and the graviton field. 
We summarize our results in section 6.
In appendix, we review the boundary states in the Green-Schwarz
formalism.

\section{Wave functions and Vertex operators}
\setcounter{equation}{0}

In this section we summarize our previous results on wave functions for a
D-instanton and vertex operators in type IIB matrix model. 
Wave functions are functions of a $d=10$ coordinate
(or its conjugate momentum) and $d=10$ Majorana Weyl spinor
and contain information on the couplings of a D-instanton to
the closed string massless states.
Their physical meaning in superstring theories will be clarified in the next
section. 
On the other hand, in the matrix model,
interactions corresponding to the supergravity modes 
are induced as quantum effects and their couplings to 
these modes are described through the vertex operators.

\subsection{Supersymmetric Wilson line operator }
The degrees of freedom of a D-instanton are described by its coordinate, a
ten-dimensional vector $y_\mu$ and a Majorana-Weyl fermion $\lambda$,
and thus information of its state is encoded in functions of $y_\mu$ and
$\lambda$, that is, wave functions. 
Here we give wave functions corresponding to the supergravity modes in the
form of $f_A(\lambda)e^{-ik\cdot y}$ with a momentum $k$, where the index
$A$ specifies each mode. 
Wave functions are defined to form a multiplet of the following
$d=10$ ${\cal N}=2$ supersymmetry transformations
\begin{eqnarray}
\delta^{(1)}f(\lambda)  &=& [\bar{\epsilon}_1q_1, f(\lambda)] = 
\epsilon_1 \frac{\partial}{\partial\lambda} f(\lambda), 
\label{susy5-1}\\
\delta^{(2)}  f(\lambda) &=& [\bar{\epsilon}_2 q_2, f(\lambda)]=
(\bar{\epsilon}_2 \kslash \lambda)  f(\lambda),
\label{susy5-2}
\end{eqnarray}
where $\epsilon_i~(i=1, 2)$ are the Majorana-Weyl spinors.

The Majorana-Weyl fermion $\lambda$ contains 16 degrees of freedom
and there are $2^{16}$ independent wave functions for $\lambda$.
To reduce the number, we impose the massless condition for the momentum
$k$; $k^2=0$. 
Then, since $\kslash \lambda $ in (\ref{susy5-2}) has only 8 independent 
degrees of freedom, the supersymmetry can generate only $2^8=256$
independent wave functions for $\lambda$. 
They form a massless type IIB supergravity multiplet
containing a complex dilaton $\Phi$, 
a complex dilatino $\tilde\Phi$, a complex antisymmetric tensor 
$B_{\mu\nu}$, a complex gravitino $\Psi_\mu$, a real graviton 
$h_{\mu\nu}$ and a real 4th-rank self-dual antisymmetric tensor
$A_{\mu\nu\rho\sigma}$. 
A physical meaning of the wave functions in string theories
is given by using boundary states of a D-instanton in section 3.

Vertex operators $V_A(A^\mu, \psi;k)$ covariantly transform under the 
following ${\cal N}=2$ supersymmetry of the IIB matrix model,
\begin{eqnarray}
 \label{supercharges}
  \left\{
   \begin{array}{l}
    \delta^{(1)} A_\mu = i \bar{\epsilon}_1 \Gamma_\mu \psi, \\
    \delta^{(1)} \psi = \displaystyle{
     -\frac{i}{2} [A_\mu , A_\nu] \Gamma^{\mu\nu} \epsilon_1,
     }
   \end{array}
      \right. 
  \qquad
  \left\{
   \begin{array}{l}
    \delta^{(2)} A_\mu = 0, \\
    \delta^{(2)} \psi = \epsilon_2 1_N.
   \end{array}
       \right.
  \label{susy}
\end{eqnarray}
We denote the generator of $\delta^{(i)}~(i=1, 2)$ as $Q_i~(i=1, 2)$,
respectively. 
Since the ${\cal N}=2$ supersymmetry algebra closes only 
on shell, in this section we assume that the $N\times N$ matrices $A_\mu$
and $\psi$ satisfy the equations of motion for the IKKT action (\ref{IKKT}),
\begin{eqnarray}
 && \label{eom-boson}
  \left[A^\nu, \left[A_\mu, A_\nu\right]\right]
  -\frac{1}{2}\left(\Gamma_0 \Gamma_\mu\right)_{\alpha\beta}
  \left\{\psi_\alpha, \psi_\beta\right\}=0, \\
 && \label{eom-fermion}
  \Gamma^\mu\left[A_\mu, \psi\right]=0.
\end{eqnarray}

In order to construct vertex operators systematically,
we start from a supersymmetric Wilson line operator first introduced 
in \cite{Hamada} for the IIB matrix model;
\be
\omega(C) = \mathrm{tr} \prod_j e^{\bar\lambda_j Q_1} 
e^{-i \epsilon k_j^\mu A_\mu} 
e^{-\bar\lambda_j Q_1}.
\ee
Since we are interested in the massless multiplet, 
we here consider the  simplest straight Wilson line operator 
with a global momentum $k$;
\be
\omega(\lambda,k)= e^{\bar\lambda Q_1}~ \mathrm{tr} e^{ik \cdot A} ~
e^{-\bar\lambda Q_1}. 
\label{omega}
\ee 
Here, though the Majorana-Weyl spinor $\lambda$ is a parameter of the
supersymmetry transformation, it is eventually interpreted as 
a fermionic collective coordinate of a D-instanton. 

This supersymmetric Wilson line operator  $\omega(\lambda,k)$ is invariant 
under simultaneous supersymmetry transformations for
$N \times N$ matrices $A^\mu, \psi$ and the parameters $(\lambda, k)$ as
\begin{eqnarray}
 && \label{susywilsoninv1}
  [ \bar{\epsilon}_1 Q_1, \omega(\lambda,k) ]-[\bar{\epsilon}_1q_1, \omega(\lambda,k) ] = 0,  \\
 && \label{susywilsoninv2}
  [ \bar{\epsilon}_2 Q_2, \omega(\lambda,k) ] -[\bar{\epsilon}_2 q_2, \omega(\lambda,k)]
  = 0.
\end{eqnarray}
By expanding $\omega(\lambda,k)$ in terms of 
the wave functions for $\lambda$, which are constructed in the manner stated
above, as
\begin{equation}
\omega(\lambda,k) = \sum_A f_A(\lambda) \ V_A(A_\mu, \psi; \ k),
\end{equation}
it is understood from eqs.(\ref{susywilsoninv1}) and (\ref{susywilsoninv2})
that $V_A(A_\mu, \psi; \ k)$ correctly transform under the ${\cal N}=2$
supersymmetry. Therefore $V_A(A_\mu, \psi; \ k)$ can be regarded as
candidates for the vertex operators. Indeed it will be shown explicitly in
section 5 that a system of $N$ D-instantons couples to the supergravity
modes through these vertex operators.

\subsection{Wave functions of a D-instanton \label{2.2}}
We here summarize our results of the wave functions for the massless
multiplet and their supersymmetry transformations. 
In constructing wave functions which transform covariantly under the
supersymmetries, we first assume that the dilaton wave function is
proportional to $\exp(-i k \cdot y)$, 
namely $f_A(\lambda)=1$. It is annihilated by the supersymmetry
transformation $q_1$. 
Then the other wave functions can be determined 
by supersymmetry transformations.
For more details, see
\cite{ITU}. 

By defining a fermion bilinear as 
$b_{\mu \nu}=k_{\sigma} \bar\lambda \Gamma_{\mu \nu \sigma} \lambda$, 
the supersymmetry multiplet of the wave functions is given as follows; 
\begin{itemize}

\item  dilaton
\be
\Phi(\lambda,k) = 1 ,
\label{dilaton1}
\ee
\item dilatino 
\be
\tilde\Phi(\lambda,k) = \kslash \lambda,
\label{dilatino1}
\ee
\item antisymmetric tensor field 
\be 
B_{\mu\nu}(\lambda,k) = - \frac{1}{2} b_{\mu\nu}(\lambda),
\label{ast1}
\ee
\item gravitino 
\be
\Psi_\mu(\lambda,k) 
= - \frac{i}{24} (k_\sigma \Gamma^{\nu\sigma} \lambda) b_{\mu\nu}(\lambda), 
\label{gravitino1}
\ee
\item graviton
\be
h_{\mu \nu}(\lambda,k) = \frac{1}{96} b_\mu^{~\rho}b_{\rho \nu}(\lambda), 
\ee
\item 4-th rank self-dual antisymmetric tensor field
\be
A_{\mu\nu\rho\sigma}(\lambda,k)=-\frac{i}{32(4!)^2}
b_{[\mu\nu}b_{\rho\sigma]}(\lambda),
\ee
\item gravitino (charge conjugation of (\ref{gravitino1}))
\be
\Psi_\mu^c(\lambda, k) =
-\frac{i}{4 \cdot 5!} k^\rho  \Gamma_{\rho \lambda}  
\lambda b^{\lambda \sigma} b_{\sigma\mu}(\lambda), 
\ee
\item antisymmetric tensor field  (charge conjugation of (\ref{ast1}))
\be
\label{ast2}
B_{\mu\nu}^c(\lambda, k) 
= -\frac{1}{6!} b_{\mu\rho}b^{\rho\sigma}b_{\sigma\nu}(\lambda), 
\ee
\item dilatino  (charge conjugation of (\ref{dilatino1}))
\be
\tilde\Phi^c(\lambda, k) =
      \frac{1}{8!}k_\alpha \Gamma^{\mu \nu \alpha} \lambda 
      b_{\nu \rho} b^{\rho \sigma} b_{\sigma \mu}(\lambda), 
\ee
\item dilaton  (charge conjugation of (\ref{dilaton1}))
\be
\Phi^c(\lambda, k) = \frac{1}{8 \cdot 8!}
b_\mu^{~ \nu} b_\nu^{~ \lambda} b_\lambda^{~ \sigma} b_\sigma^{~ \mu}(\lambda).
\label{dilaton2}
\ee
\end{itemize}
In these expressions we have chosen a specific 
gauge for each wave function.
These wave functions can be interpreted as  overlaps of D-instanton boundary 
states and closed string massless states as we will see  in the next  section.
In the usual convention of superstrings, the first dilaton (\ref{dilaton1}) 
corresponds to a wave function of (dilaton $+i$ axion) and the second one
(\ref{dilaton2}) corresponds to (dilaton $-i$ axion). 
The other complex fields also have the same structure.

The supersymmetry transformations (\ref{susy5-1}, \ref{susy5-2}) lead to the
following transformations between these wave functions;

\bea
 \label{susy-transformation}
\delta \Phi &=& \bar\epsilon_2 \tilde\Phi, \nonumber \\
\delta \tilde\Phi &=& \kslash \epsilon_1 \Phi -\frac{i}{24} 
\Gamma^{\mu\nu\rho} \epsilon_2 H_{\mu\nu\rho}, \nonumber \\
\delta B_{\mu\nu} &=& - \bar\epsilon_1 \Gamma_{\mu\nu}\tilde\Phi 
+ 2i(\bar\epsilon_2 \Gamma_{[\mu}\Psi_{\nu]} + k_{[\mu}\Lambda_{\nu]}), 
\nonumber \\
\delta \Psi_\mu 
&=& \frac{1}{24\cdot4} \left[ 9\Gamma^{\nu \rho}\epsilon_1 H_{\mu\nu\rho} 
-\Gamma_{\mu\nu\rho\sigma}\epsilon_1 H^{\nu\rho\sigma} \right]
+\frac{i}{2} \Gamma^{\nu\rho} k_\rho h_{\mu \nu}\epsilon_2
\nonumber \\
&& + \frac{i}{4 \cdot 5!} \Gamma^{\rho_1 \cdots \rho_5} 
\Gamma_\mu \epsilon_2 F_{\rho_1 \cdots \rho_5} + k_\mu \xi, 
\nonumber\\
\delta h_{\mu\nu} &=& -\frac{i}{2} \bar\epsilon_1\Gamma_{(\mu}\Psi_{\nu)}
-\frac{i}{2} \bar\epsilon_2\Gamma_{(\mu}\Psi^c_{\nu)} 
+ k_{(\mu} \xi_{\nu)}, 
\nonumber \\
\delta A_{\mu\nu\rho\sigma} 
&=& 
-\frac{1}{(4!)^2} \bar\epsilon_1 \Gamma_{[\mu\nu\rho}\Psi_{\sigma]} 
-\frac{1}{(4!)^2} \bar\epsilon_2 \Gamma_{[\mu\nu\rho}\Psi^c_{\sigma]}
+ k_{[\mu} \xi_{\nu\rho\sigma]},
\nonumber \\
\delta \Psi_\mu^c
&=& \frac{i}{2} \Gamma^{\nu\rho} k_\rho h_{\mu \nu}\epsilon_1
+ \frac{i}{4 \cdot 5!} \Gamma^{\rho_1 \cdots \rho_5} 
\Gamma_\mu\epsilon_1 F_{\rho_1 \cdots \rho_5}
\nonumber \\
&& +\frac{1}{24\cdot4} \left[ 9\Gamma^{\nu \rho}\epsilon_2 H_{\mu\nu\rho}^c 
-{\Gamma_\mu}^{\nu\rho\sigma}\epsilon_2 H_{\nu\rho\sigma}^c \right]
+ k_\mu \xi^c, 
\nonumber \\
\delta B_{\mu\nu}^c 
&=& 2i(\bar\epsilon_1 \Gamma_{[\mu}\Psi_{\nu]}^c + k_{[\mu}\Lambda_{\nu]}^c)  
 -\bar\epsilon_2 \Gamma_{\mu\nu} \tilde\Phi^c,
\nonumber \\ 
\delta\tilde\Phi^c 
&=& - \frac{i}{24} \Gamma^{\mu\nu\rho} \epsilon_1 H_{\mu\nu\rho}^c 
+ \kslash\epsilon_2 \Phi^c, \nonumber \\
\delta \Phi^c &=& \bar\epsilon_1 \tilde\Phi^c,
\label{susytransformations}
\eea
where $\xi, \xi^c, \xi_\mu, \xi_{\mu\nu\rho}, \Lambda_\mu$ 
and $\Lambda^c_\mu$ are gauge parameters. 
$H_{\mu\nu\rho}, H_{\mu\nu\rho}^c$ and $F_{\rho_1 \cdots \rho_5}$ are 
the field strengths of $B_{\mu\nu}, B_{\mu\nu}^c$ and $A_{\mu\nu\rho\sigma}$, 
respectively.
This supersymmetry transformation is the same as that in \cite{Schwarz:1983wa}
up to  normalizations.

\subsection{Vertex operators}

Construction of the vertex operators can be done systematically by expanding 
the supersymmetric Wilson line operator in terms of the wave functions 
$f_A(\lambda)$ given in the previous subsection. 
In section 5, we will show that these vertex operators indeed describe
couplings of  
type IIB matrix model to the supergravity modes.
The derivation itself is systematic but the complete calculation is cumbersome.
Partial results were obtained in \cite{Kitazawa:2002vh}. 
More complete analysis was given in \cite{ITU}
\footnote{Similar calculations were performed in the BFSS matrix model
in \cite{Taylor-1} and \cite{Dasgupta:2000df}.}.
The results are as follows. 
\begin{itemize}
\item dilaton 
\be
V^\Phi =\mathrm{tr}~ e^{ik \cdot A},
\ee
\item dilatino 
\be
V^{\tilde\Phi} = \mathrm{tr}~e^{ik \cdot A}\bar\psi,
\ee
\item antisymmetric tensor field 
\bea
V^{B}_{\mu \nu} 
&=& \mathrm{Str}~e^{ik \cdot A}
\left( 
\frac{1}{16}k^\rho(\bar\psi \cdot \Gamma_{\mu\nu\rho}\psi) 
-\frac{i}{2}[A_\mu,A_\nu] 
\right),
\eea
\item gravitino 
\bea
V^{\Psi}_{\mu} 
&=& 
\mathrm{Str}~e^{ik\cdot A} 
\left(
-\frac{i}{12}k^\rho(\bar\psi \cdot \Gamma_{\mu\nu\rho} \psi) 
-2[A_\mu, A_\nu] 
\right)
\cdot \bar{\psi}\Gamma^\nu,
\eea

\item graviton 
\bea
\label{graviton-vertex}
V^h_{\mu\nu}
&=&
2~\mathrm{Str}~e^{ik\cdot A}
\biggl\{
[A_\mu,A^\rho] \cdot [A_\nu,A_\rho] 
+ \frac{1}{4}\bar\psi \cdot \Gamma_{(\mu}[A_{\nu)},\psi]
-\frac{i}{8}k^\rho 
\bar\psi \cdot \Gamma_{\rho\sigma(\mu}\psi \cdot [A_{\nu)}, A^\sigma]
 \nonumber \\
&& \hspace{19mm}
-\frac{1}{8\cdot4!} k^\lambda k^\tau 
(\bar\psi \cdot \Gamma_{\mu\lambda}^{~~~\sigma}\psi)
\cdot (\bar\psi \cdot \Gamma_{\nu\tau\sigma}\psi)
\biggl\},
\eea
\item 4-th rank self-dual antisymmetric tensor field
\bea
 V^A_{\mu\nu\rho\sigma} 
  &=&
  -i~\mathrm{Str}~e^{ik\cdot A}
  \biggl\{
  F_{[\mu\nu} \cdot F_{\rho\sigma]} 
  +c\bar{\psi} \cdot \Gamma_{[\mu\nu\rho}[A_{\sigma]}, \psi]
  - \frac{3i}{4}c k^\lambda
  \bar\psi \cdot \Gamma_{\lambda[\mu\nu}\psi \cdot F_{\rho\sigma]}
  \nonumber \\
 && \hspace{35mm}
  -\frac{1}{8\cdot 4!} k^\lambda k^\tau 
  (\bar\psi \cdot \Gamma_{\lambda[\mu\nu}\psi)
  \cdot (\bar\psi \cdot \Gamma_{\rho\sigma]\tau}\psi)
  \biggr\},
  \label{4tensorvertexop}
\eea
where $c=-1/3$. We fixed the value of $c$ by another calculation 
(See Section IV-E in \cite{ITU}). 
\end{itemize}

Hereafter we write down only the leading order terms of vertex operators.
\begin{itemize}
\item charge conjugation of gravitino
\be
 V_\mu^{\Psi^c} = 
   \mathrm{Str}~ e^{ik\cdot A} 
   \left( [A_\mu, A_\nu]\cdot [A_\rho, A_\sigma]\cdot 
    \Gamma^{\rho\sigma}\Gamma^{\nu} \psi
    +\frac{2}{3}\bar{\psi}\cdot \Gamma_{\nu}[A_\mu, \psi]\cdot
    \Gamma^{\nu} \psi
   \right),
\ee
\item charge conjugation of antisymmetric tensor field
\be
      \label{Vbc}
 \mathrm{Str}~ e^{ik\cdot A}
  \left(
  [A_\mu, A_\rho]\cdot [A^\rho, A^\sigma]\cdot [A_\sigma, A_\nu]
  -\frac{1}{4}[A_\mu, A_\nu]\cdot [A^\rho, A^\sigma]\cdot [A_\sigma, A_\rho]
  \right),
\ee
\item charge conjugation of dilatino
\bea
 V^{\tilde{\Phi}^c} &=&
  \mathrm{Str}~e^{ik\cdot A}
  \left\{
  \left(
  [A_\mu, A_\rho]\cdot [A^\rho, A^\sigma]\cdot [A_\sigma, A_\nu]
  -\frac{1}{4}[A_\mu, A_\nu]\cdot [A^\rho, A^\sigma]
  \cdot [A_\sigma, A_\rho]
  \right)
  \cdot \Gamma^{\mu\nu}\psi
  \right. \nonumber \\
  && \left.
  +\frac{1}{24}[A_\mu, A_\nu]\cdot [A_\rho, A_\sigma]
  \cdot [A_\lambda, A_\tau]
  \cdot \Gamma^{\mu\nu\rho\sigma\lambda\tau}\psi
  \right\},
\eea
\item charge conjugation of dilaton
\bea
  V^{\Phi^c} &=& 
  \mathrm{Str}~ e^{ik\cdot A}
  \bigg\{
   [A_\mu, A_\nu]\cdot [A^\nu, A^\rho]\cdot [A_\rho, A_\sigma]
   \cdot [A^\sigma, A^\mu]
    \nonumber \\
   && 
   -\frac{1}{4}[A_\mu, A_\nu] \cdot[A^\nu, A^\mu]
   \cdot [A_\rho, A_\sigma]\cdot [A^\sigma, A^\rho]
   \nonumber \\
   && 
   +  [A_\sigma,  A_\mu]   \cdot [A_\nu, A_\rho]  
      \cdot \bar\psi \Gamma^\mu \Gamma^{\nu \rho} \cdot  [A_\sigma, \psi]
   \bigg\}.
   \label{ccdilaton}
\eea

\end{itemize}
$\mathrm{Str}$ means a symmetrized  trace which is defined by
\begin{eqnarray}
 \label{def-str}
 \mathrm{Str}~ e^{ik\cdot A} B_1\cdot B_2 \cdots B_n
  &=& 
  \int_0^1 dt_1 \int_{t_1}^1 dt_2 \cdots \int_{t_{n-2}}^1 dt_{n-1}
  \nonumber \\
 && 
  \times \mathrm{tr}~ e^{ik\cdot A t_1} B_1 e^{ik\cdot A(t_2-t_1)} B_2 
  \cdots 
  e^{ik\cdot A(t_{n-1}-t_{n-2})} B_{n-1} 
  e^{ik\cdot A(1-t_{n-1})} B_n
  \nonumber \\
 &&
  + \left(\mbox{ permutations of  $B_i$'s} \ (i=2, 3, \cdots, n)~\right).
\end{eqnarray}
The center-dot on the left hand side means that the operators $B_i$ 
are symmetrized. 
In the first term in (\ref{graviton-vertex}), for example, 
$B_1$ and $B_2$ correspond to $[A_\mu, A^\rho]$ and $[A_\nu, A_\rho]$, 
respectively.
See the appendix of \cite{ITU} for various properties of the symmetrized trace.
For notational simplicity we sometimes use $\mathrm{Str}$ also for a single
operator like
$
 \mathrm{Str}~ (e^{i k \cdot A} B)
$
which is equivalent to the ordinary trace.

\section{Stringy Interpretation of Wave Functions}
\setcounter{equation}{0}

In this section we show that the wave functions obtained in the previous
section can be interpreted as overlaps of D-instanton boundary states and
closed string massless states in the Green-Schwarz formalism of type IIB
superstring. 
The ordinary D-instanton is known to be coupled only with the dilaton and
the axion states~\cite{Green} and 
becomes a source for these closed string modes only.
But the D-instanton is a half-BPS state and breaks a half of the
supersymmetries and we can construct a supersymmetry multiplet by acting 
broken supersymmetry generators successively on the simplest D-instanton
boundary state. 
Namely the D-instanton has an internal structure and
these multiplet states are coupled also to the other closed string massless
states such as gravitons or antisymmetric tensor fields. 
Hence they become a source for these fields, although the 
couplings contain higher derivatives. 
Such internal structures of D-branes were discussed in various papers
\cite{Green-Gutperle-1, Green-Gutperle-2, Morales-1, Duff, Morales-2,
Taylor-1, Taylor-2, Millar}.
In the following, we show that the wave functions in the previous section
are nothing but overlaps of such D-instanton boundary states with the
closed string massless states.

We adopt the Green-Schwarz formalism of type IIB superstring and
take the light-cone gauge. 
Our notations and brief summaries of a construction of boundary states in the
Green-Schwarz formalism are given in the appendix. 
Definitions of the supercharges and a boundary state for the D-instanton are 
obtained by setting 
\begin{equation}
 \label{setting}
 \eta = +1, \quad M_{ij}=\delta_{ij}, \quad M_{ab}=\delta_{ab}, \quad 
  M_{\dot{a}\dot{b}}=\delta_{\dot{a}\dot{b}},
\end{equation}
in the corresponding equations in the appendix 
((\ref{bc-1A})-(\ref{broken-charge2})).

The type IIB superstring has  ${\cal N}=2$ supersymmetries with 32 
supercharges.
A boundary state for the D-instanton is defined by 
the boundary conditions in eqs.(\ref{bc-1A})-(\ref{bc-3A}) with 
(\ref{setting}),
\begin{eqnarray}
 && \partial_\sigma X^i |B\rangle = 0, 
 \nonumber \\
 && Q^{+a} |B \rangle = 0, 
  \nonumber \\
 && Q^{+\dot{a}} |B \rangle = 0,
  \label{bc-3}
\end{eqnarray}
and a solution of these conditions is given in eq.(\ref{boundary-state}) as
\begin{equation}
 |B \rangle =
  e^{\sum_{n>0}
  \left(
   \frac{1}{n}\alpha^i_{-n}\tilde{\alpha}^i_{-n}
   -i S^a_{-n}\tilde{S}^a_{-n}
  \right)}
  |B_0 \rangle,
\end{equation} 
where $S^a_n$ and $\tilde{S}^a_n$ are fermionic modes and $\alpha^i_n$ and
$\tilde{\alpha}^i_n$ are bosonic modes of the type IIB superstring.
From eq.(\ref{bs-zero}), the zero-mode part becomes 
\begin{equation}
 |B_0  \rangle 
  = C\left( |i\rangle |i\rangle
     -i |\dot{a}\rangle |\dot{a}\rangle\right),
\end{equation}
where $C$ is a normalization constant.
The D-instanton boundary state preserves a half of supersymmetries 
$Q^{+a}$ and $Q^{+\dot{a}}$, and breaks the other half $Q^{-a}$ and
$Q^{-\dot{a}}$ which are defined in (\ref{broken-charge1}) and
(\ref{broken-charge2}). 
The broken and unbroken supercharges satisfy the algebra
\begin{eqnarray}
 \label{bualgebra}
 \{Q^{+a}, Q^{-b}\} &=& 4p^+ \delta_{ab}, \nonumber \\
 \{Q^{+a}, Q^{-\dot{b}}\} &=& 
  2\sqrt{2}\gamma^i_{a\dot{b}}p^i, \nonumber \\
 \{Q^{+\dot{a}}, Q^{-b}\} &=&
  2\sqrt{2}\gamma^i_{\dot{a}b}p^i, \nonumber \\
 \{Q^{+\dot{a}}, Q^{-\dot{b}}\} &=&
  2(P^- + \tilde{P}^-) \delta_{\dot{a}\dot{b}}.
\end{eqnarray}
The other anticommutators vanish.
\subsection{Coupling of D-instanton boundary states to supergravity modes}
States  obtained by acting the broken generators $Q^{-a}, Q^{-\dot{a}}$ 
on the D-instanton boundary states couple to the supergravity modes.
Here we concentrate on massless modes and ignore massive excitations.

The zero-mode part of the boundary state of the D-instanton is  given by
\begin{equation}
 |D(-1)\rangle =\frac{1}{\sqrt{2}}\left( |i\rangle |i\rangle 
  -i |\dot{a}\rangle |\dot{a}\rangle\right), 
\end{equation}
where we set the normalization constant $C=1/\sqrt{2}$ for simplicity.
This state couples to  a linear combination of the dilaton and axion,
\begin{equation}
 |\Phi \rangle \equiv 
  \frac{1}{\sqrt{2}}( |i\rangle |i\rangle 
  -i |\dot{a}\rangle |\dot{a}\rangle ).
\end{equation}
The coupling is given by 
\begin{equation}
 \langle \Phi |D(-1)\rangle = 1.
\end{equation}

Acting the broken charge $\lambda^a Q^{-a}$ on $|D(-1)\rangle$, 
we obtain the fermionic state 
\begin{equation}
 \lambda^a Q^{-a} |D(-1)\rangle 
  = \sqrt{2p^+} \gamma^i_{a\dot{a}}\lambda^a
  \left(|\dot{a}\rangle |i\rangle 
   -i|i\rangle |\dot{a}\rangle \right).
\end{equation}
This couples to the following linear combination of dilatino states
\begin{equation}
 |\tilde{\Phi}_a \rangle \sim 
  \sqrt{p^+}\gamma^i_{a\dot{a}}\left(|\dot{a}\rangle |i\rangle  
   -i|i\rangle |\dot{a}\rangle \right),
\end{equation}
and the coupling is given by
\begin{equation}
 \langle \tilde{\Phi}_a | \lambda^b Q^{-b} |D(-1)\rangle 
  \sim p^+ \lambda^a.
\end{equation}
The normalizations of states for the supergravity modes are fixed so that 
the supersymmetry transformations of them satisfy 
eqs.(\ref{susytransformations}). 

By further acting the broken supersymmetry charges, we can construct
the following state
\begin{eqnarray}
 &&
 \lambda^{a_1}\lambda^{a_2}Q^{-a_1}Q^{-a_2}|D(-1)\rangle
 \nonumber \\
 && \hspace{5mm}
  = 2\sqrt{2}p^+ \gamma^{ij}_{a_1a_2}\lambda^{a_1}\lambda^{a_2}
  |i\rangle |j\rangle
  -\sqrt{2}ip^+ \left(\gamma^i_{a_1\dot{a}}\gamma^i_{a_2\dot{b}}
	 -\gamma^i_{a_2\dot{a}}\gamma^i_{a_1\dot{b}}\right)
  \lambda^{a_1}\lambda^{a_2}
  |\dot{a}\rangle |\dot{b}\rangle.
\label{qqboundary}
\end{eqnarray}
This state couples to the antisymmetric tensor field $B_{\mu\nu}$,
\begin{eqnarray}
 |B_{ij}\rangle \sim
  |i\rangle |j\rangle - |j\rangle |i\rangle
  -\frac{i}{2}\gamma^{ij}_{\dot{a}\dot{b}}|\dot{a}\rangle |\dot{b}\rangle.
\end{eqnarray}
The coupling between these states is given by
\begin{equation}
 \langle B_{ij}|\lambda^{a_1}\lambda^{a_2}Q^{-a_1}Q^{-a_2}|D(-1)\rangle
  \sim p^+ 
  \left(\gamma^{ij}_{a_1a_2}\lambda^{a_1}\lambda^{a_2}\right).
\end{equation}
Since the coupling contains momentum $p^+$, the boundary state 
(\ref{qqboundary}) has a derivative-coupling to the antisymmetric 
tensor field.

The state multiplied by three broken charges is given by
\begin{eqnarray}
 &&
  \lambda^{a_1}\lambda^{a_2}\lambda^{a_3}
  Q^{-a_1}Q^{-a_2}Q^{-a_3} |D(-1)\rangle 
  \nonumber \\
 && \hspace{5mm}
  = \left(2p^+\right)^{\frac{3}{2}}
  \left[
   \gamma^j_{a_1\dot{a}}\gamma^{ji}_{a_2a_3}
   +\frac{1}{2}\gamma^i_{a_1\dot{b}}
   \left(\gamma^j_{a_2\dot{a}}\gamma^j_{a_3\dot{b}}
   - \gamma^j_{a_3\dot{a}}\gamma^j_{a_2\dot{b}}\right)
  \right]
  \lambda^{a_1}\lambda^{a_2}\lambda^{a_3}
  \left(|\dot{a}\rangle |i\rangle +i |i\rangle |\dot{a}\rangle\right).
  \nonumber \\
 \label{qqqboundary} 
\end{eqnarray} 
This state couples to a linear combination of gravitino states 
\begin{eqnarray}
 |\Psi^{\dot{a}}_i \rangle 
  \sim \sqrt{p^+} 
  \left[
   |\dot{a}\rangle |i\rangle + i |i\rangle |\dot{a}\rangle
   -\frac{1}{8}\gamma^i_{\dot{a}b}\gamma^j_{b\dot{b}}
   \left(|\dot{b}\rangle |j\rangle + i |j\rangle |\dot{b}\rangle \right)
  \right].
\label{gravitinostate}
\end{eqnarray}
Hence the coupling between the boundary state (\ref{qqqboundary}) and the
gravitino state (\ref{gravitinostate}) becomes 
\begin{eqnarray}
 \langle \Psi^{\dot{a}}_i |\lambda^{a_1}\lambda^{a_2}\lambda^{a_3}
  Q^{-a_1}Q^{-a_2}Q^{-a_3} |D(-1)\rangle 
  \sim \left(p^+\right)^2
  \gamma^j_{\dot{a}a_1}\lambda^{a_1}
  \left(\gamma^{ji}_{a_2a_3}\lambda^{a_2}\lambda^{a_3}\right).
\end{eqnarray}

A boundary state which is obtained by acting  four broken generators on
$|D(-1)\rangle$ becomes
\begin{eqnarray}
 &&\lambda^{a_1}\cdots\lambda^{a_4}
  Q^{-a_1}\cdots Q^{-a_4} |D(-1)\rangle
  \nonumber \\
 &&
  = 8\sqrt{2} (p^+)^2 
  \left[
   \left(\gamma^{ik}_{a_1a_2}\lambda^{a_1}\lambda^{a_2}\right)
   \left(\gamma^{kj}_{a_3a_4}\lambda^{a_3}\lambda^{a_4}\right)
   |i\rangle |j\rangle 
   -i\left(\gamma^{ij}_{a_3a_4}\lambda^{a_3}\lambda^{a_4}\right)
   \left(\gamma^i_{a_1\dot{a}}\gamma^j_{a_2\dot{b}}
    \lambda^{a_1}\lambda^{a_2}\right)
   |\dot{a}\rangle |\dot{b}\rangle
  \right].
  \nonumber \\
\end{eqnarray}
This state couples to the graviton state
\begin{equation}
 |h_{ij}\rangle 
  \sim |i\rangle |j\rangle + |i\rangle |j\rangle
  -\frac{1}{4}\delta_{ij}|k\rangle|k\rangle,
\end{equation}
and its coupling is given by
\begin{eqnarray}
 \langle h_{ij}|\lambda^{a_1}\cdots\lambda^{a_4}
  Q^{-a_1}\cdots Q^{-a_4} |D(-1)\rangle 
  \sim \left(p^+\right)^2
  \left(\gamma^{ik}_{a_1a_2}\lambda^{a_1}\lambda^{a_2}\right)
   \left(\gamma^{kj}_{a_3a_4}\lambda^{a_3}\lambda^{a_4}\right).
\end{eqnarray}
The coupling contains two derivatives.

We can similarly construct states by acting 
more broken supersymmetry generators.
They couple to the other massless states of type IIB 
closed string through derivative couplings\footnote{
Note that both of $Q^{-a}$ and $Q^{-\dot{a}}$ have the same structure 
$S^a_0 -i \tilde{S}^a_0$ in the zero-mode part.  Hence as long as the
massless closed string states are concerned, it is sufficient to consider only 
one of those two generators, namely $\lambda^a Q^{-a}$.
}.

\subsection{Wave functions with light-cone momentum}
In order to compare the wave functions of the mean-field D-instanton in
section \ref{2.2} to the results in the previous subsection,
we take the light-cone momentum and rewrite the wave functions
in section \ref{2.2}.
 
Let us take the frame where the momentum is represented as 
\begin{equation}
 \label{lc-momentum}
 k^\mu = \left(E, 0, \cdots , 0, E\right),
\end{equation}
namely, only the $k^+$ component is non-vanishing.
Then the following relations hold:
\begin{eqnarray*}
 && \kslash = E\left(\Gamma_0 + \Gamma_9\right) 
  = -E \left(\Gamma^0 - \Gamma^9\right)
  = -\sqrt{2}E \Gamma^{-}\\
 && \kslash \lambda = 2iE\left(\lambda^a, 0, -\lambda^a, 0\right)^T, \\
 && b_{ij} = 4E \left(\gamma^{ij}_{ab}\lambda^a \lambda^b\right), 
  \qquad b_{i-} = 4\sqrt{2}E
  \left(\gamma^i_{\dot{a}a}\lambda^{\dot{a}}\lambda^a\right), \\
 && b_{i+} = 0.
\end{eqnarray*}
By using these relations, transverse components
of the wave functions 
in section \ref{2.2} become
\begin{eqnarray}
 \Phi &=& 1, \nonumber \\
 \tilde{\Phi} &=& 2iE\left(\lambda^a, 0, -\lambda^a, 0\right)^T, 
  \qquad \left(\Gamma_{11} \tilde{\Phi}= + \tilde{\Phi}\right)
  \nonumber \\
 B_{ij} &=& -2E\left(\gamma^{ij}_{ab}\lambda^a\lambda^b\right), 
  \nonumber \\
 \Psi_i &=& 4iE^2 \left(\gamma^{ij}_{bc}\lambda^b\lambda^c\right)
  \left(0, -\gamma^j_{\dot{a}a}\lambda^a, 0, \gamma^j_{\dot{a}a}\lambda^a
  \right)^T, \nonumber \\
  h_{ij} &=& \frac{1}{6}E^2 
  \left(\gamma^{ik}_{ab}\lambda^a\lambda^b\right)
  \left(\gamma^{kj}_{cd}\lambda^c\lambda^d\right), \nonumber \\
\end{eqnarray}
They are the same as the overlaps in the previous subsection 
with the identification $p^+ = \sqrt{2}E$, up to normalizations.
Hence we have shown that the wave functions of the mean-field D-instanton
represent couplings of a supersymmetry multiplet of a D-instanton to closed
string massless states.

\subsection{Fermionic coherent state of D-instanton}
So far we have constructed boundary states by acting a fixed number of
broken supersymmetry generators on $|D(-1) \rangle$ so that they form an
ordinary set of a supersymmetry multiplet. 
In order to see the above interpretation more systematically, we construct a 
fermionic coherent state by acting the unitary operator 
$\exp(-\lambda^a Q^{-a})$ on $|D(-1) \rangle$; 
\be
|\lambda \rangle = \exp(-\lambda^a Q^{-a}) |D(-1) \rangle.
\ee 
Due to the commutation relations (\ref{bualgebra}), this state satisfies 
modified boundary conditions 
\bea
\label{Q-coh1}
Q^{+a} |\lambda \rangle &=& 4p^+ \lambda^a |\lambda \rangle  \\
Q^{+\dot{a}}|\lambda \rangle &=& 2\sqrt{2}
p^i\gamma^i_{\dot{a}a}\lambda^a
 |\lambda\rangle.
 \label{Q-coh2}
\eea
In the IIB matrix model, the bosonic coordinates are interpreted as
the coordinates of space-time. From the consideration here, the fermionic
coordinates can be interpreted as the fermionic parameters which bestow an
internal structure on the space-time constructed from the bosonic
coordinates. 

The wave functions for the mean-field D-instanton can be written as 
\begin{eqnarray}
 \label{wf-overlap}
 f_A(\lambda) = \langle A |\lambda \rangle,
\end{eqnarray}
for each supergravity state $A$.
In the previous subsection, this relation has been shown separately for each
state $|A\rangle$ up to normalization.
It can be understood more directly as follows.
When the momentum $k$ is taken as (\ref{lc-momentum}), 
the supercharges $q_1$ and $q_2$ for a D-instanton, (\ref{susy5-1}) and
(\ref{susy5-2}), have the following forms,
\begin{eqnarray}
 q_1^a = -i\frac{\partial}{\partial\lambda^a}, \qquad
  q_2^a = 2iE \lambda^a,
\end{eqnarray} 
and satisfy the algebra
\begin{equation}
\{q_1^a, \ q_2^b\} = 2E\delta_{ab}, \qquad \mbox{others} = 0.
\end{equation}
On the other hand, as far as  massless states are concerned, this algebra
is equivalent to the ones among the supercharges $Q^{\pm a}$ and
$Q^{\pm\dot{a}}$ with $p^i=P^-=\tilde{P}^-=0$, eq.(\ref{bualgebra}).
Actually actions of $q_i^a$ on the wave functions (\ref{wf-overlap}) can be 
regarded as insertions  of $Q^{-a}$ and $Q^{+a}$ which act on the
massless state of the supergravity modes $|A\rangle$ as follows,
\begin{eqnarray*}
 &&
  q_1^a f_A(\lambda) = -i\frac{\partial}{\partial\lambda}f_A(\lambda)
  = i\left(\langle A |Q^{-a}\right)|\lambda\rangle, \\
 &&
  q_2^a f_A(\lambda) = 2iE\lambda^a f_A(\lambda)
  = \frac{i}{2\sqrt{2}}\left(\langle A|Q^{+a}\right)|\lambda\rangle,
\end{eqnarray*}
where we have used eqs.(\ref{Q-coh1}) and (\ref{Q-coh2}).
Hence a construction of the supergravity multiplet by acting $Q^{\pm a}$ 
on the closed string massless state $\langle A|$
corresponds to the one by acting $q_i^a$ on wave functions $f_A(\lambda)$ 
and
 the wave functions we constructed describe the (derivative)
couplings between a D-instanton and various supergravity modes.

\section{One-loop Effective Action}
\setcounter{equation}{0}

In the latter half of the paper, we discuss condensation of massless
supergravity fields in type IIB matrix model. We consider a
matrix model of size $(N+1) \times (N+1)$ 
and integrate over one D-instanton with the wave functions given
in section 2. 
In this way, we can obtain a modified effective action in a weak supergravity
background of $N$ D-instantons.

In this section we first give a systematic evaluation of the one-loop
effective action with general fermionic backgrounds.
The results were partly given in \cite{suyama} and \cite{Kimura-Kitazawa}.
Similar calculations were performed in the BFSS matrix model in 
\cite{Taylor-1}.
Since we are interested in condensation, we do not use
the matrix model equation of motion in this section.

We start from the type IIB matrix model with a size $(N+1)\times(N+1)$
and write $(N+1)\times (N+1)$ bosonic and fermionic hermitian
matrices as
$A'_\mu \ (\mu=0, \cdots, 9)$ and $\psi'$.  
We then decompose them  into backgrounds $\left(X_\mu, \Phi\right)$
and fluctuations $\left(a_\mu, \varphi\right)$ around them as
\begin{eqnarray}
 \label{decomposition}
 A'_\mu &=& X_\mu + a_\mu, \nonumber \\
 \psi' &=& \Phi + \varphi. 
\end{eqnarray}
In order to perform perturbative calculations, we fix a gauge and add the
following terms to the action (\ref{IKKT}),
\begin{equation}
 \label{gf}
 S_{g.f. + ghost} = -\mathrm{tr}
  \left(
   \frac{1}{2}\left[X^\mu, a_\mu\right]\left[X^\nu, a_\nu\right]
   +\left[X_\mu, b\right]\left[A'^\mu, c\right]
  \right),
\end{equation}
where $c$ and $b$ are ghost and anti-ghost fields respectively.
Substituting the decompositions (\ref{decomposition}) into the action
(\ref{IKKT}) and (\ref{gf}), we obtain the following expression up to the 2nd
order of the fluctuations,
\begin{eqnarray}
 S_{IKKT}+ S_{g.f. + ghost} &=& S_{IKKT}(X, \Phi)
  -\frac{1}{2}\mathrm{tr}
  \left(
   a^\mu\left[X^\nu, \left[X_\nu,a_\mu \right]\right]
   +2a_\mu\left[\left[X^\mu, X^\nu\right], a_\nu\right]
  \right) \nonumber \\
 &&
  -\frac{1}{2}\mathrm{tr}~\bar{\varphi}\Gamma^\mu\left[X_\mu, \varphi\right] 
  -\mathrm{tr}~\bar{\Phi}\Gamma^\mu\left[a_\mu, \varphi\right]
  +\mathrm{tr}~b\left[X_\mu, \left[X^\mu, c\right]\right]
  \nonumber \\
 &=& S_{IKKT}(X, \Phi)+\frac{1}{2}\mathrm{tr}~
  a_\mu\left(\delta_{\mu\nu}\tilde{X}^2 + 2\tilde{F}_{\mu\nu}
       + \tilde{\bar{\Phi}}\Gamma_\mu\frac{1}{\Gamma\cdot\tilde{X}}\Gamma_\nu
       \tilde{\Phi}\right)a_\nu \nonumber \\
 && -\frac{1}{2}\mathrm{tr}~
  \left(
   \bar{\varphi}+\left[\bar{\Phi}, a_\mu\right]\Gamma_\mu
   \frac{1}{\Gamma\cdot\tilde{X}}
  \right)
  \left(\Gamma\cdot\tilde{X}\right)
  \left(
   \varphi+\frac{1}{\Gamma\cdot\tilde{X}}\Gamma_\nu\left[a_\nu, \Phi\right]
  \right)
  \nonumber \\
 && + \mathrm{tr}~b\tilde{X}^2 c + \mbox{higher orders},
\end{eqnarray}
where we defined $F_{\mu\nu}=\left[X_\mu, X_\nu\right]$,
$\Gamma\cdot X\equiv\Gamma_\mu X_\mu$
and denoted the adjoint action of a general operator $O$ as 
$\tilde{O}S\equiv\left[O, S\right]$. 
Then the one-loop partition function of the IIB matrix model becomes 
\begin{eqnarray}
 Z(X, \Phi) &=& 
  \int da_\mu d\varphi db dc~ 
  e^{-\left(S_{IKKT} + S_{g.f. + ghost}\right)}
  \nonumber \\
 &\sim& 
  e^{-S_{IKKT}(X, \Phi)}
  det^{-\frac{1}{2}}
  \left(
   \delta_{\mu\nu}\tilde{X}^2 + 2\tilde{F}_{\mu\nu}
  +\tilde{\bar{\Phi}}\Gamma_\mu\frac{1}{\Gamma\cdot\tilde{X}}
  \Gamma_\nu\tilde{\Phi}
  \right)
  \nonumber \\
 && \times
  det^{\frac{1}{4}}
  \left(
   \left(
    \tilde{X}^2+\frac{1}{2}\Gamma_{\mu\nu}\tilde{F}_{\mu\nu}
   \right)
   \frac{1+\Gamma_{11}}{2}
  \right)
  det \left(\tilde{X}^2\right).
\end{eqnarray}
Thus the free energy is given by
\begin{eqnarray}
 \label{free energy}
 F(X, \Phi) &=& -\ln Z(X, \Phi) = S_{IKKT}(X, \Phi) + F_b + F_f, \\ 
 F_b &=& \frac{1}{2}{\cal T}r \ln 
  \left(\delta_{\mu\nu}\tilde{X}^2+2\tilde{F}_{\mu\nu}\right)
  -\frac{1}{4}{\cal T}r \ln
  \left(
   \tilde{X}^2+\frac{1}{2}\Gamma_{\mu\nu}\tilde{F}_{\mu\nu}
   \frac{1+\Gamma_{11}}{2}
  \right)
  \nonumber \\
 &&
  -{\cal T}r \ln \tilde{X}^2, \\
 F_f &=&\frac{1}{2}{\cal T}r \ln
  \left[
   \delta_{\mu\nu} + \left(\frac{1}{\tilde{X}^2+2\tilde{F}}\right)_{\mu\rho} 
   \tilde{\bar{\Phi}}\Gamma_{\rho}\frac{1}{\Gamma\cdot\tilde{X}}
   \Gamma_\nu\tilde{\Phi}
  \right],
\end{eqnarray}
where ${\cal T}r$ is the trace of the adjoint operators.

We first expand $F_f$ formally with respect to the inverse powers of
$\tilde{X}$. 
To this end we use the following formulas,
\begin{eqnarray}
 \frac{1}{\tilde{X}^2+2\tilde{F}} 
  &=&
  \frac{1}{1+\frac{2}{\tilde{X}^2}\tilde{F}} \frac{1}{\tilde{X}^2}, 
  \label{boson-propagator}\\
 \frac{1}{\Gamma\cdot \tilde{X}} 
  &=&
  \frac{1}{1+\frac{1}{2\tilde{X}^2}\Gamma\cdot\tilde{F}}\frac{1}{\tilde{X}^2}
  \Gamma\cdot \tilde{X} 
  \label{fermion-propagator1}\\
 &=& 
  \frac{1}{2}\frac{1}{1+\frac{1}{2\tilde{X}^2}\Gamma\cdot\tilde{F}}
  \frac{1}{\tilde{X}^2}\Gamma\cdot \tilde{X}
  +\frac{1}{2}\Gamma\cdot \tilde{X}
  \frac{1}{1+\frac{1}{2\tilde{X}^2}\Gamma\cdot\tilde{F}}
  \frac{1}{\tilde{X}^2},
  \label{fermion-propagator2}
\end{eqnarray} 
where 
\begin{equation}
 \left(\Gamma\cdot \tilde{X}\right)^2
  = \tilde{X}^2 + \frac{1}{2}\Gamma\cdot\tilde{F},
\end{equation}
and $\Gamma\cdot \tilde{F} \equiv \Gamma_{\mu\nu}\tilde{F}_{\mu\nu}$.
In the following we expand the free energy with respect to $1/\tilde{X}$.
Since the leading part of $\tilde{X}$ is a distance between $N$ D-instantons
and a single D-instanton, this expansion is valid when the single D-instanton
is far separated from the other $N$ D-instantons. 
 
\subsection{Second order terms of $\mathbf{\Phi}$}
First let us focus on the terms with two fermions.
As is seen in section 5, these terms are relevant for condensation of the
antisymmetric tensor field.
After using eqs.(\ref{boson-propagator}) and (\ref{fermion-propagator2}),
the second order terms of the fermionic background $\Phi$ are given by
\begin{eqnarray}
 F_f \Big|_{\Phi^2}
  &=& \frac{1}{4}{\cal T}r 
  \left[
   \left(\frac{1}{1+\frac{2}{\tilde{X}^2}\tilde{F}}\right)_{\mu\nu}
   \frac{1}{\tilde{X}^2}\tilde{\bar{\Phi}}\Gamma_\nu
   \frac{1}{1+\frac{1}{2\tilde{X}^2}\Gamma\cdot\tilde{F}}
   \frac{1}{\tilde{X}^2}
   \left(\Gamma\cdot\tilde{X}\right)
   \Gamma_\mu\tilde{\Phi}
   \right. \nonumber \\
 && \left. \hspace{10mm}
     +\left(\frac{1}{1+\frac{2}{\tilde{X}^2}\tilde{F}}\right)_{\mu\nu}
     \frac{1}{\tilde{X}^2}\tilde{\bar{\Phi}}\Gamma_\nu
     \left(\Gamma\cdot\tilde{X}\right)
     \frac{1}{1+\frac{1}{2\tilde{X}^2}\Gamma\cdot\tilde{F}}
     \frac{1}{\tilde{X}^2}\Gamma_\mu\tilde{\Phi}
    \right].
    \label{phiphi}
\end{eqnarray}
We now expand the effective action with two $\Phi$'s (\ref{phiphi})
with respect to $1/\tilde{X}$.

\subsubsection{$\tilde{X}^{-3}$}
The leading order starts from $1/\tilde{X}^3$
and is given by 
\begin{eqnarray}
 \label{phi^2-x^{-3}}
 &&
  \frac{1}{4}{\cal T}r 
  \left[
   \frac{1}{\tilde{X}^2}\tilde{\bar{\Phi}}\frac{1}{\tilde{X}^2}
   \Gamma_\mu \left(\Gamma\cdot\tilde{X}\right)\Gamma_\mu\tilde{\Phi}
   +\frac{1}{\tilde{X}^2}\tilde{\bar{\Phi}}\Gamma_\mu
   \left(\Gamma\cdot\tilde{X}\right)\Gamma_\mu
   \frac{1}{\tilde{X}^2}\tilde{\Phi}
  \right] 
  \nonumber \\
 && = -2{\cal T}r~ \frac{1}{\tilde{X}^2}\tilde{\bar{\Phi}}
  \frac{1}{\tilde{X}^2}\Gamma_\mu
  \left[\tilde{X}_\mu, \tilde{\Phi}\right].
\end{eqnarray}
This is proportional to the equation of motion for the fermion.

\subsubsection{$\tilde{X}^{-5}$}
The next-to-leading order is proportional to $1/\tilde{X}^5$.
At this order we have the following terms,
\begin{eqnarray}
 &&\frac{1}{4}{\cal T}r
  \left[
   \left(-\frac{2}{\tilde{X}^2}\tilde{F}_{\mu\nu}\right)
   \frac{1}{\tilde{X}^2}\tilde{\bar{\Phi}}\Gamma_\nu
   \frac{1}{\tilde{X}^2}\left(\Gamma\cdot\tilde{X}\right)
   \Gamma_\mu\tilde{\Phi}
   +\frac{1}{\tilde{X}^2}\tilde{\bar{\Phi}}\Gamma_\mu
   \left(-\frac{1}{2\tilde{X}^2}\Gamma\cdot\tilde{F}\right)
   \frac{1}{\tilde{X}^2}\left(\Gamma\cdot\tilde{X}\right)
   \Gamma_\mu\tilde{\Phi}
  \right. 
  \nonumber \\
 && \left. \hspace{5mm}
  +\left(-\frac{2}{\tilde{X}^2}\tilde{F}_{\mu\nu}\right)
  \frac{1}{\tilde{X}^2}\tilde{\bar{\Phi}}\Gamma_\nu
  \left(\Gamma\cdot\tilde{X}\right)
  \frac{1}{\tilde{X}^2}\Gamma_\mu\tilde{\Phi}
  +\frac{1}{\tilde{X}^2}\tilde{\bar{\Phi}}\Gamma_\mu
  \left(\Gamma\cdot\tilde{X}\right)
  \left(-\frac{1}{2\tilde{X}^2}\Gamma\cdot\tilde{F}\right)
  \frac{1}{\tilde{X}^2}\Gamma_\mu\tilde{\Phi}
  \right].
  \nonumber \\
\end{eqnarray}
After some calculations, these terms are rewritten as
\begin{eqnarray}
 \label{phi^2-x^{-5}}
 &&-\frac{1}{2}{\cal T}r~
  \frac{1}{\tilde{X}^2}\tilde{F}_{\mu\nu}
  \frac{1}{\tilde{X}^2}\tilde{\bar{\Phi}}
  \frac{1}{\tilde{X}^2}\Gamma_{\mu\nu}\cdot
  \Gamma_\rho\left[\tilde{X}_\rho, \tilde{\Phi}\right]
  -\frac{1}{2}{\cal T}r~
  \frac{1}{\tilde{X}^2}\tilde{F}_{\mu\nu}
  \frac{1}{\tilde{X}^2}
  \left[\tilde{\bar{\Phi}}, \tilde{X}_\rho\right]
  \Gamma_\rho\cdot\Gamma_{\mu\nu}
  \frac{1}{\tilde{X}^2}\tilde{\Phi} 
  \nonumber \\
 && -{\cal T}r~
  \frac{1}{\tilde{X}^2}\tilde{F}_{\mu\nu}
  \frac{1}{\tilde{X}^2}\tilde{\bar{\Phi}}
  \frac{1}{\tilde{X}^2}\Gamma_\nu
  \tilde{X}_\mu\tilde{\Phi}
  -{\cal T}r~
  \frac{1}{\tilde{X}^2}\tilde{F}_{\mu\nu}
  \frac{1}{\tilde{X}^2}\tilde{\bar{\Phi}}
  \frac{1}{\tilde{X}^2}\Gamma_\nu
  \tilde{\Phi}\tilde{X}_\mu
  \nonumber \\
 && +{\cal T}r~
  \frac{1}{\tilde{X}^2}\tilde{F}_{\mu\nu}
  \frac{1}{\tilde{X}^2}\tilde{X}_\mu
  \tilde{\bar{\Phi}}
  \frac{1}{\tilde{X}^2}\Gamma_\nu
  \tilde{\Phi}
  +{\cal T}r~
  \frac{1}{\tilde{X}^2}\tilde{F}_{\mu\nu}
  \frac{1}{\tilde{X}^2}
  \tilde{\bar{\Phi}}\tilde{X}_\mu
  \frac{1}{\tilde{X}^2}\Gamma_\nu
  \tilde{\Phi}.
\end{eqnarray}
The first two terms are proportional to the equation of motion.
It is noted that the terms in the second and the third lines vanish if
$\tilde{X}_\mu$ is replaced with $d_\mu$. 
Here $d_{\mu}$ is a vector directed to the center of the $N$
D-instantons from the single D-instanton.
Therefore these terms are actually ${\cal O}(d^{-6})$ in the $1/d$
expansions. 

\subsubsection{$\tilde{X}^{-7}$}
The terms of the order $\tilde{X}^{-7}$ are given by 
\begin{eqnarray}
 && \hspace{-10mm}
  \frac{1}{4}{\cal T}r
  \left[
   \left(\frac{2}{\tilde{X}^2}\tilde{F}_{\mu\nu}\right)
   \left(\frac{2}{\tilde{X}^2}\tilde{F}_{\nu\rho}\right)
   \frac{1}{\tilde{X}^2}
   \tilde{\bar{\Phi}}\Gamma_\rho\frac{1}{\tilde{X}^2}
   \left(\Gamma\cdot\tilde{X}\right)
   \Gamma_\mu\tilde{\Phi} 
   \right. 
  \nonumber \\
 &&
  +\frac{1}{\tilde{X}^2}\tilde{\bar{\Phi}}\Gamma_\mu
  \left(\frac{1}{2\tilde{X}^2}\Gamma\cdot\tilde{F}\right)
  \left(\frac{1}{2\tilde{X}^2}\Gamma\cdot\tilde{F}\right)
  \frac{1}{\tilde{X}^2}
  \left(\Gamma\cdot\tilde{X}\right)
  \Gamma_\mu\tilde{\Phi}
  \nonumber \\
 &&
  +\left(\frac{2}{\tilde{X}^2}\tilde{F}_{\mu\nu}\right)
  \frac{1}{\tilde{X}^2}
  \tilde{\bar{\Phi}}\Gamma_\nu
  \left(\frac{1}{2\tilde{X}^2}\Gamma\cdot\tilde{F}\right)
  \frac{1}{\tilde{X}^2}
  \left(\Gamma\cdot\tilde{X}\right)
  \Gamma_\mu\tilde{\Phi}
  \nonumber \\
 &&
  +\left(\frac{2}{\tilde{X}^2}\tilde{F}_{\mu\nu}\right)
  \left(\frac{2}{\tilde{X}^2}\tilde{F}_{\nu\rho}\right)
  \frac{1}{\tilde{X}^2}
  \tilde{\bar{\Phi}}\Gamma_\rho
  \left(\Gamma\cdot\tilde{X}\right)
  \frac{1}{\tilde{X}^2}
  \Gamma_\mu\tilde{\Phi}
  \nonumber \\
 &&
  +\frac{1}{\tilde{X}^2}\tilde{\bar{\Phi}}\Gamma_\mu
  \left(\Gamma\cdot\tilde{X}\right)
  \left(\frac{1}{2\tilde{X}^2}\Gamma\cdot\tilde{F}\right)
  \left(\frac{1}{2\tilde{X}^2}\Gamma\cdot\tilde{F}\right)
  \frac{1}{\tilde{X}^2}
  \Gamma_\mu\tilde{\Phi}
  \nonumber \\
 && 
  \left.
   +\left(\frac{2}{\tilde{X}^2}\tilde{F}_{\mu\nu}\right)
   \frac{1}{\tilde{X}^2}
   \tilde{\bar{\Phi}}\Gamma_\nu
   \left(\Gamma\cdot\tilde{X}\right)
   \left(\frac{1}{2\tilde{X}^2}\Gamma\cdot\tilde{F}\right)
   \frac{1}{\tilde{X}^2}
   \Gamma_\mu\tilde{\Phi}
  \right].
\end{eqnarray}
These are rewritten as
\begin{eqnarray}
 \label{phi^2-x^{-7}}
 &&
  \frac{1}{4}{\cal T}r~
  \frac{1}{\tilde{X}^2}\tilde{F}_{\mu\nu}\frac{1}{\tilde{X}^2}
  \tilde{\bar{\Phi}}
  \frac{1}{\tilde{X}^2}\tilde{F}_{\rho\sigma}\frac{1}{\tilde{X}^2}
  \Gamma_{\mu\nu\rho\sigma}\Gamma_\lambda
  \left[\tilde{X}_\lambda, \tilde{\Phi}\right]
  \nonumber \\
 &&
  +{\cal T}r~
  \frac{1}{\tilde{X}^2}\tilde{F}_{\mu\nu}\frac{1}{\tilde{X}^2}
  \tilde{F}_{\nu\rho}\frac{1}{\tilde{X}^2}
  \tilde{\bar{\Phi}}
  \frac{1}{\tilde{X}^2}
  \Gamma_{\mu\rho}\Gamma_\sigma
  \left[\tilde{X}_\sigma, \tilde{\Phi}\right]
  +{\cal T}r~
  \frac{1}{\tilde{X}^2}\tilde{F}_{\mu\nu}\frac{1}{\tilde{X}^2}
  \tilde{F}_{\nu\rho}\frac{1}{\tilde{X}^2}
  \left[\tilde{\bar{\Phi}}, \tilde{X}_\sigma\right]
  \Gamma_\sigma\Gamma_{\mu\rho}
  \frac{1}{\tilde{X}^2}
  \tilde{\Phi}
  \nonumber \\
 &&
  -{\cal T}r~
  \frac{1}{\tilde{X}^2}\tilde{F}_{\mu\nu}\frac{1}{\tilde{X}^2}
  \tilde{F}_{\nu\mu}\frac{1}{\tilde{X}^2}
  \tilde{\bar{\Phi}}
  \frac{1}{\tilde{X}^2}
  \Gamma_\rho
  \left[\tilde{X}_\rho, \tilde{\Phi}\right]
  -{\cal T}r~
  \frac{1}{\tilde{X}^2}\tilde{F}_{\mu\nu}\frac{1}{\tilde{X}^2}
  \tilde{F}_{\nu\mu}\frac{1}{\tilde{X}^2}
  \left[\tilde{\bar{\Phi}}, \tilde{X}_\rho\right]
  \Gamma_\rho
  \frac{1}{\tilde{X}^2}
  \tilde{\Phi}
  \nonumber \\
 &&
  -\frac{1}{2}{\cal T}r~
  \frac{1}{\tilde{X}^2}\tilde{F}_{\mu\nu}\frac{1}{\tilde{X}^2}
  \tilde{\bar{\Phi}}
  \frac{1}{\tilde{X}^2}\tilde{F}_{\nu\mu}\frac{1}{\tilde{X}^2}
  \Gamma_\rho
  \left[\tilde{X}_\rho, \tilde{\Phi}\right]
  \nonumber \\
 &&
  -\frac{1}{2}{\cal T}r~
  \frac{1}{\tilde{X}^2}\tilde{F}_{\mu\nu}\frac{1}{\tilde{X}^2}
  \tilde{F}_{\rho\sigma}\frac{1}{\tilde{X}^2}
  \tilde{X}_\nu
  \tilde{\bar{\Phi}}
  \Gamma_{\mu\rho\sigma}
  \frac{1}{\tilde{X}^2}
  \tilde{\Phi}
  -\frac{1}{2}{\cal T}r~
  \frac{1}{\tilde{X}^2}\tilde{F}_{\mu\nu}\frac{1}{\tilde{X}^2}
  \tilde{F}_{\rho\sigma}\frac{1}{\tilde{X}^2}
  \tilde{X}_\sigma
  \tilde{\bar{\Phi}}
  \Gamma_{\mu\nu\rho}
  \frac{1}{\tilde{X}^2}
  \tilde{\Phi}
  \nonumber \\
 &&
  -\frac{1}{2}{\cal T}r~
  \frac{1}{\tilde{X}^2}\tilde{F}_{\mu\nu}\frac{1}{\tilde{X}^2}
  \tilde{F}_{\rho\sigma}\frac{1}{\tilde{X}^2}
  \tilde{\bar{\Phi}}
  \Gamma_{\mu\nu\sigma}
  \frac{1}{\tilde{X}^2}
  \tilde{\Phi}
  \tilde{X}_\rho
  -\frac{1}{2}{\cal T}r~
  \frac{1}{\tilde{X}^2}\tilde{F}_{\mu\nu}\frac{1}{\tilde{X}^2}
  \tilde{F}_{\rho\sigma}\frac{1}{\tilde{X}^2}
  \tilde{\bar{\Phi}}
  \Gamma_{\nu\rho\sigma}
  \frac{1}{\tilde{X}^2}
  \tilde{\Phi}
  \tilde{X}_\mu
  \nonumber \\
 &&
  -\frac{1}{2}{\cal T}r~
  \frac{1}{\tilde{X}^2}\tilde{F}_{\mu\nu}\frac{1}{\tilde{X}^2}
  \tilde{\bar{\Phi}}
  \frac{1}{\tilde{X}^2}\tilde{F}_{\rho\sigma}\frac{1}{\tilde{X}^2}
  \tilde{X}_\nu
  \Gamma_{\mu\rho\sigma}
  \tilde{\Phi}
  +\frac{1}{2}{\cal T}r~
  \frac{1}{\tilde{X}^2}\tilde{F}_{\mu\nu}\frac{1}{\tilde{X}^2}
  \tilde{\bar{\Phi}}
  \tilde{X}_\nu
  \frac{1}{\tilde{X}^2}\tilde{F}_{\rho\sigma}\frac{1}{\tilde{X}^2}
  \Gamma_{\mu\rho\sigma}
  \tilde{\Phi}
  \nonumber \\
 &&
  +2{\cal T}r~
  \frac{1}{\tilde{X}^2}\tilde{F}_{\mu\nu}\frac{1}{\tilde{X}^2}
  \tilde{F}_{\nu\rho}\frac{1}{\tilde{X}^2}
  \tilde{\bar{\Phi}}
  \frac{1}{\tilde{X}^2}
  \Gamma_\rho
  \tilde{X}_\mu
  \tilde{\Phi}
  +2{\cal T}r~
  \frac{1}{\tilde{X}^2}\tilde{F}_{\mu\nu}\frac{1}{\tilde{X}^2}
  \tilde{F}_{\nu\rho}\frac{1}{\tilde{X}^2}
  \tilde{\bar{\Phi}}
  \tilde{X}_\rho
  \frac{1}{\tilde{X}^2}
  \Gamma_\mu
  \tilde{\Phi}
  \nonumber \\
 &&
  -{\cal T}r~
  \frac{1}{\tilde{X}^2}\tilde{F}_{\mu\nu}\frac{1}{\tilde{X}^2}
  \tilde{F}_{\nu\rho}\frac{1}{\tilde{X}^2}
  \tilde{\bar{\Phi}}
  \Gamma_\mu
  \frac{1}{\tilde{X}^2}
  \tilde{\Phi}
  \tilde{X}_\rho
  +{\cal T}r~
  \frac{1}{\tilde{X}^2}\tilde{F}_{\mu\nu}\frac{1}{\tilde{X}^2}
  \tilde{F}_{\nu\rho}\frac{1}{\tilde{X}^2}
  \tilde{\bar{\Phi}}
  \Gamma_\rho
  \frac{1}{\tilde{X}^2}
  \tilde{\Phi}
  \tilde{X}_\mu
  \nonumber \\
 &&
  -{\cal T}r~
  \frac{1}{\tilde{X}^2}\tilde{F}_{\mu\nu}\frac{1}{\tilde{X}^2}
  \tilde{F}_{\nu\rho}\frac{1}{\tilde{X}^2}
  \tilde{X}_\mu
  \tilde{\bar{\Phi}}
  \Gamma_\rho
  \frac{1}{\tilde{X}^2}
  \tilde{\Phi}
  +{\cal T}r~
  \frac{1}{\tilde{X}^2}\tilde{F}_{\mu\nu}\frac{1}{\tilde{X}^2}
  \tilde{F}_{\nu\rho}\frac{1}{\tilde{X}^2}
  \tilde{X}_\rho
  \tilde{\bar{\Phi}}
  \Gamma_\mu
  \frac{1}{\tilde{X}^2}
  \tilde{\Phi}
  \nonumber \\
 &&
  +{\cal T}r~
  \frac{1}{\tilde{X}^2}\tilde{F}_{\mu\nu}\frac{1}{\tilde{X}^2}
  \tilde{\bar{\Phi}}
  \frac{1}{\tilde{X}^2}\tilde{F}_{\nu\rho}\frac{1}{\tilde{X}^2}
  \tilde{X}_\mu
  \Gamma_\rho
  \tilde{\Phi}
  +{\cal T}r~
  \frac{1}{\tilde{X}^2}\tilde{F}_{\mu\nu}\frac{1}{\tilde{X}^2}
  \tilde{\bar{\Phi}}
  \tilde{X}_\mu
  \frac{1}{\tilde{X}^2}\tilde{F}_{\nu\rho}\frac{1}{\tilde{X}^2}
  \Gamma_\rho
  \tilde{\Phi}.
\end{eqnarray}
The first six terms vanish if the fermionic background satisfies the
equation of motion.

\subsubsection{$\tilde{X}^{-9}$}
The terms of the order $\tilde{X}^{-9}$ are given by 
\begin{eqnarray}
 \label{phi^2-x^{-9}}
 {\cal T}r~
  \Bigg[ \hspace{-5mm} &&
  -2\frac{1}{\tilde{X}^2}\tilde{F}_{\mu\nu}\frac{1}{\tilde{X}^2}
  \tilde{F}_{\nu\rho}\frac{1}{\tilde{X}^2}
  \tilde{F}_{\rho\sigma}\frac{1}{\tilde{X}^2}
  \tilde{\bar{\Phi}}\Gamma_\sigma \frac{1}{\tilde{X}^2}
  \left(\Gamma\cdot \tilde{X}\right)
  \Gamma_\mu \tilde{\Phi}  \nonumber \\
 &&
  -\frac{1}{32}\frac{1}{\tilde{X}^2}\tilde{\bar{\Phi}}
  \Gamma_\mu \frac{1}{\tilde{X}^2}
  \left(\Gamma\cdot\tilde{F}\right)\frac{1}{\tilde{X}^2}
  \left(\Gamma\cdot\tilde{F}\right)\frac{1}{\tilde{X}^2}
  \left(\Gamma\cdot\tilde{F}\right)\frac{1}{\tilde{X}^2}
  \left(\Gamma\cdot \tilde{X}\right)
  \Gamma_\mu \tilde{\Phi} \nonumber \\
 &&
  -\frac{1}{2}\frac{1}{\tilde{X}^2}\tilde{F}_{\mu\nu}\frac{1}{\tilde{X}^2}
   \tilde{F}_{\nu\rho}\frac{1}{\tilde{X}^2}
   \tilde{\bar{\Phi}}\Gamma_\rho\frac{1}{\tilde{X}^2}
   \left(\Gamma\cdot\tilde{F}\right)\frac{1}{\tilde{X}^2}
   \left(\Gamma\cdot \tilde{X}\right)
   \Gamma_\mu \tilde{\Phi} \nonumber \\
 &&
  -\frac{1}{8}\frac{1}{\tilde{X}^2}\tilde{F}_{\mu\nu}\frac{1}{\tilde{X}^2}
  \tilde{\bar{\Phi}}\Gamma_\nu\frac{1}{\tilde{X}^2}
  \left(\Gamma\cdot\tilde{F}\right)\frac{1}{\tilde{X}^2}
  \left(\Gamma\cdot\tilde{F}\right)\frac{1}{\tilde{X}^2}
  \left(\Gamma\cdot \tilde{X}\right)
   \Gamma_\mu \tilde{\Phi} \nonumber \\
 &&
  -2\frac{1}{\tilde{X}^2}\tilde{F}_{\mu\nu}\frac{1}{\tilde{X}^2}
  \tilde{F}_{\nu\rho}\frac{1}{\tilde{X}^2}
  \tilde{F}_{\rho\sigma}\frac{1}{\tilde{X}^2}
  \tilde{\bar{\Phi}}\Gamma_\sigma 
  \left(\Gamma\cdot \tilde{X}\right)
  \frac{1}{\tilde{X}^2}
  \Gamma_\mu \tilde{\Phi}  \nonumber \\
 &&
  -\frac{1}{32}\frac{1}{\tilde{X}^2}\tilde{\bar{\Phi}}
  \Gamma_\mu 
  \left(\Gamma\cdot \tilde{X}\right)  
  \frac{1}{\tilde{X}^2}
  \left(\Gamma\cdot\tilde{F}\right)\frac{1}{\tilde{X}^2}
  \left(\Gamma\cdot\tilde{F}\right)\frac{1}{\tilde{X}^2}
  \left(\Gamma\cdot\tilde{F}\right)\frac{1}{\tilde{X}^2}
  \Gamma_\mu \tilde{\Phi} \nonumber \\
 &&
  -\frac{1}{2}\frac{1}{\tilde{X}^2}\tilde{F}_{\mu\nu}\frac{1}{\tilde{X}^2}
   \tilde{F}_{\nu\rho}\frac{1}{\tilde{X}^2}
   \tilde{\bar{\Phi}}\Gamma_\rho
   \left(\Gamma\cdot \tilde{X}\right)
   \frac{1}{\tilde{X}^2}
   \left(\Gamma\cdot\tilde{F}\right)\frac{1}{\tilde{X}^2}
   \Gamma_\mu \tilde{\Phi} \nonumber \\
 &&
  -\frac{1}{8}\frac{1}{\tilde{X}^2}\tilde{F}_{\mu\nu}\frac{1}{\tilde{X}^2}
  \tilde{\bar{\Phi}}\Gamma_\nu
  \left(\Gamma\cdot \tilde{X}\right)
  \frac{1}{\tilde{X}^2}
  \left(\Gamma\cdot\tilde{F}\right)\frac{1}{\tilde{X}^2}
  \left(\Gamma\cdot\tilde{F}\right)\frac{1}{\tilde{X}^2}
  \Gamma_\mu \tilde{\Phi} 
  \Bigg].
\end{eqnarray}
\subsection{Fourth order terms of $\mathbf{\Phi}$}
Now let us consider four-fermion terms.
These terms are relevant for condensation of gravitons.
The fourth order terms of $\Phi$ are given by
\begin{eqnarray}
 F_f\Big|_{\Phi^4} 
  &=&
  -\frac{1}{4}{\cal T}r
  \left(\frac{1}{1+\frac{2}{\tilde{X}^2}\tilde{F}}\right)_{\mu\nu}
  \frac{1}{\tilde{X}^2}
  \tilde{\bar{\Phi}}\Gamma_\nu
  \frac{1}{1+\frac{1}{2\tilde{X}^2}\Gamma\cdot \tilde{F}}
  \frac{1}{\tilde{X}^2}
  \left(\Gamma\cdot\tilde{X}\right)
  \Gamma_\rho\tilde{\Phi} 
  \nonumber \\
 && \hspace{10mm} \times
  \left(\frac{1}{1+\frac{2}{\tilde{X}^2}\tilde{F}}\right)_{\rho\sigma}
  \frac{1}{\tilde{X}^2}
  \tilde{\bar{\Phi}}\Gamma_\sigma
  \frac{1}{1+\frac{1}{2\tilde{X}^2}\Gamma\cdot \tilde{F}}
  \frac{1}{\tilde{X}^2}
  \left(\Gamma\cdot\tilde{X}\right)
  \Gamma_\mu\tilde{\Phi}.
  \label{phi4}
\end{eqnarray}  

\subsubsection{$\tilde{X}^{-6}$}
The leading order term is proportional to $1/\tilde{X}^6$.
The term of this order is given by 
\begin{equation}
 \label{phi^4-x^{-6}}
 -\frac{1}{4}{\cal T}r~
  \frac{1}{\tilde{X}^2}\tilde{\bar{\Phi}}
  \Gamma_\mu \frac{1}{\tilde{X}^2}
  \left(\Gamma\cdot\tilde{X}\right)
  \Gamma_\nu \tilde{\Phi}
  \frac{1}{\tilde{X}^2}\tilde{\bar{\Phi}}
  \Gamma_\nu \frac{1}{\tilde{X}^2}
  \left(\Gamma\cdot\tilde{X}\right)
  \Gamma_\mu \tilde{\Phi}.
\end{equation}

\subsubsection{$\tilde{X}^{-8}$}
The next order terms are proportional to $1/\tilde{X}^8$ and given by
\begin{eqnarray}
 &&
  \label{phi^4-x^{-8}}
  -\frac{1}{2}{\cal T}r
  \left[
   \left(-\frac{2}{\tilde{X}^2}\tilde{F}_{\mu\nu}\right)
   \frac{1}{\tilde{X}^2}
   \tilde{\bar{\Phi}} \Gamma_\nu \frac{1}{\tilde{X}^2}
   \left(\Gamma\cdot \tilde{X}\right)
   \Gamma_\rho \tilde{\Phi} \frac{1}{\tilde{X}^2}
   \tilde{\bar{\Phi}}\Gamma_\rho \frac{1}{\tilde{X}^2}
   \left(\Gamma\cdot \tilde{X}\right)
   \Gamma_\mu \tilde{\Phi}
  \right.
  \nonumber \\
 && \hspace{10mm} 
  \left.
   +\frac{1}{\tilde{X}^2}
   \tilde{\bar{\Phi}} \Gamma_\mu 
   \left(-\frac{1}{2\tilde{X}^2}\Gamma\cdot\tilde{F}\right)
   \frac{1}{\tilde{X}^2}   
   \left(\Gamma\cdot \tilde{X}\right)
   \Gamma_\nu \tilde{\Phi} \frac{1}{\tilde{X}^2}
   \tilde{\bar{\Phi}}\Gamma_\nu \frac{1}{\tilde{X}^2}
   \left(\Gamma\cdot \tilde{X}\right)
   \Gamma_\mu \tilde{\Phi}
  \right].
\end{eqnarray}

\subsubsection{$\tilde{X}^{-10}$}
The terms of the order $1/\tilde{X}^{10}$ become
\begin{eqnarray}
 \label{phi^4-x^{-10}}
 && \hspace{-5mm}
  -{\cal T}r
  \left[
   2\frac{1}{\tilde{X}^2}\tilde{F}_{\mu\nu}\frac{1}{\tilde{X}^2}
   \tilde{F}_{\nu\rho}\frac{1}{\tilde{X}^2}
   \tilde{\bar{\Phi}}\Gamma_{\rho}
   \frac{1}{\tilde{X}^2}\left(\Gamma\cdot \tilde{X}\right)
   \Gamma_\sigma \tilde{\Phi}
   \frac{1}{\tilde{X}^2}\tilde{\bar{\Phi}}
   \Gamma_\sigma\frac{1}{\tilde{X}^2}\left(\Gamma\cdot \tilde{X}\right)
   \Gamma_\mu \tilde{\Phi}
   \right. \nonumber \\
 && \hspace{5mm}
  +\frac{1}{8}\frac{1}{\tilde{X}^2}\tilde{\bar{\Phi}}\Gamma_\mu
  \frac{1}{\tilde{X}^2}\Gamma_{\lambda\rho}\tilde{F}_{\lambda\rho}
  \frac{1}{\tilde{X}^2}\Gamma_{\sigma\tau}\tilde{F}_{\sigma\tau}
  \frac{1}{\tilde{X}^2}\left(\Gamma\cdot \tilde{X}\right)
  \Gamma_\nu\tilde{\Phi}\frac{1}{\tilde{X}^2}
  \tilde{\bar{\Phi}}\Gamma_\nu\frac{1}{\tilde{X}^2}
  \left(\Gamma\cdot \tilde{X}\right)
  \Gamma_\mu \tilde{\Phi}
  \nonumber \\
 && \hspace{5mm}
  +\frac{1}{2}\frac{1}{\tilde{X}^2}
  F_{\mu\nu}\frac{1}{\tilde{X}^2}\tilde{\bar{\Phi}}
  \Gamma_\nu \frac{1}{\tilde{X}^2}
  \Gamma_{\sigma\tau} \tilde{F}_{\sigma\tau}\frac{1}{\tilde{X}^2}
  \left(\Gamma\cdot \tilde{X}\right)
  \Gamma_\rho \tilde{\Phi}\frac{1}{\tilde{X}^2}
  \tilde{\bar{\Phi}}\Gamma_\rho
  \frac{1}{\tilde{X}^2}
  \left(\Gamma\cdot \tilde{X}\right)\Gamma_\mu \tilde{\Phi}
  \nonumber \\
 && \hspace{5mm}
  +\frac{1}{2}\frac{1}{\tilde{X}^2}F_{\mu\nu}\frac{1}{\tilde{X}^2}
  \tilde{\bar{\Phi}}\Gamma_\nu\frac{1}{\tilde{X}^2}
  \left(\Gamma\cdot \tilde{X}\right)
  \Gamma_\rho \tilde{\Phi}\frac{1}{\tilde{X}^2}
  \tilde{\bar{\Phi}}\Gamma_\rho\frac{1}{\tilde{X}^2}
  \Gamma_{\sigma\tau}\tilde{F}_{\sigma\tau}\frac{1}{\tilde{X}^2}
  \left(\Gamma\cdot \tilde{X}\right)
  \Gamma_\mu \tilde{\Phi}
  \nonumber \\
 && \hspace{5mm}
  +\frac{1}{\tilde{X}^2}F_{\mu\nu}\frac{1}{\tilde{X}^2}
  \tilde{\bar{\Phi}}\Gamma_\nu \frac{1}{\tilde{X}^2}
  \left(\Gamma\cdot \tilde{X}\right)
  \Gamma_\rho \tilde{\Phi}
  \frac{1}{\tilde{X}^2}F_{\rho\sigma}\frac{1}{\tilde{X}^2}
  \tilde{\bar{\Phi}}\Gamma_\sigma
  \frac{1}{\tilde{X}^2}\left(\Gamma\cdot \tilde{X}\right)
  \Gamma_\mu \tilde{\Phi}
  \nonumber \\
 && \hspace{5mm} \left.
  +\frac{1}{16}\tilde{\bar{\Phi}}\Gamma_\mu
  \frac{1}{\tilde{X}^2}\Gamma_{\lambda\rho}\tilde{F}_{\lambda\rho}
  \frac{1}{\tilde{X}^2}\left(\Gamma\cdot \tilde{X}\right)
  \Gamma_\nu \tilde{\Phi}\frac{1}{\tilde{X}^2}
  \tilde{\bar{\Phi}}\Gamma_\nu\frac{1}{\tilde{X}^2}
  \Gamma_{\sigma\tau} \tilde{F}_{\sigma\tau}\frac{1}{\tilde{X}^2}
  \left(\Gamma\cdot \tilde{X}\right)
  \Gamma_\mu \tilde{\Phi}
  \right].
\end{eqnarray}

\section{Condensation of the Supergravity Modes}
\setcounter{equation}{0}
In this section, we discuss modifications of the effective
actions for the IIB matrix model by condensation of D-instantons
with appropriate wave functions. They 
correspond to condensation of massless type IIB supergravity fields.
We here consider backgrounds produced by a mean-field D-instanton.
As we saw in sections 2 and 3, a D-instanton forms a supersymmetry
multiplet by acting broken supersymmetry generators on the ordinary
D-instanton boundary state. These states couple to the 
closed string massless states through derivative couplings and become a
source for these fields. 

Now we write $(N+1) \times (N+1)$ matrices in the decomposition
(\ref{decomposition}) as follows, 
\begin{eqnarray}
 &&
  X_\mu = 
  \left(
   \begin{array}{cc}
    x_\mu 1_N + A_\mu & 0 \\
    0 & y_\mu
   \end{array}
  \right), 
  \qquad
  \Phi = 
  \left(
   \begin{array}{cc}
    \psi & 0 \\
    0 & \xi
   \end{array}
  \right), 
  \\
 &&
  a_\mu =
  \left(
   \begin{array}{cc}
    0 & \alpha_\mu \\
    \alpha_\mu^\dagger & 0
   \end{array}
  \right),
  \qquad
  \varphi =
  \left(
   \begin{array}{cc}
    0 & \phi \\
    \phi^\dagger & 0
   \end{array}
  \right).
\end{eqnarray} 
$A_\mu$ and $\psi$ are $N\times N$  traceless matrices. 
$y_\mu$ is a bosonic coordinate of a (mean-field) D-instanton.
$\xi$ is a fermionic coordinate and a Majorana-Weyl spinor.
They represent degrees of freedom of the mean-field D-instanton.
Off-diagonal components $\alpha_\mu$ and $\phi$ are $N$-vectors 
corresponding to interactions between the diagonal blocks, which we have
integrated out at the one-loop order in the previous section.
Hence  the free energy (\ref{free energy}) is a function of these diagonal
components, $F(X, \Phi)=F(A, x, \psi; y, \xi)$. 
By choosing wave functions $f_k(y, \xi)$ for the mean-field D-instanton
and integrating over $y, \xi$,
we can obtain a modified effective action $S_{eff}(A, x, \psi; f_k)$ by
condensation of the massless modes; 
\begin{equation}
 e^{-S_{eff}(A, x, \psi; f_k)}
  = \int dy d\xi~ e^{-F(A, x, \psi; y, \xi)} f_k(y, \xi).
\end{equation}

In what follows, 
we mainly look at terms without fermionic matrices $\psi$
and replace all fermionic variables by the D-instanton 
fermionic coordinate $\xi.$

\subsection{Condensation of the dilaton}
In order to express condensation of dilaton in terms
of the wave function of the mean-field D-instanton,
we put $f_k (y, \xi)$ as
\be
f_{D}(y, \xi) 
= \int d^{10}k \  e^{ik\cdot y} \ \tilde{f}_D(k, \xi) 
= \int d^{10}k \  e^{ik\cdot y} \
f(k) \   \left(\prod_{\gamma =1}^{16}\xi_\gamma\right) . 
\label{dilaton16}
\ee
Then the $\xi$ integration is already saturated 
by the wave function. 
The leading contribution to  the effective action 
is easily shown to be proportional to 
(the charge conjugation of ) the
dilaton vertex operator 
(\ref{ccdilaton}).
\subsection{Condensation of the antisymmetric tensor $\bf{B_{\mu\nu}}$}
We now calculate the effective action with an insertion of a wave function
describing the antisymmetric tensor field  $B_{\mu\nu}$. 
In the present calculation,  
the supersymmetry multiplet starts from the
dilaton wave function (\ref{dilaton16})
and the other functions in the multiplet can be
constructed by acting a derivative operator $\partial/\partial \xi$.
 
By replacing $\lambda$ in eq.(\ref{ast1}) with
$\partial/\partial \xi$ and applying the
differential operator on $ \left(\prod_{\gamma =1}^{16}\xi_\gamma\right)$,
we obtain the wave function for the antisymmetric tensor field;
\begin{eqnarray}
 \label{wavefunction-B}
 f_B(y, \xi) 
 &=& \int d^{10}k \ e^{ik\cdot y} \ \tilde{f}_B(k, \xi) \\ \nonumber
 &=& \int d^{10}k
 \  e^{ik\cdot y} \ 
  ( \zeta_{\mu\nu}(k)k_\rho +\zeta_{\nu\rho}(k)k_\mu +\zeta_{\rho\mu}(k)k_\nu ) \\ \nonumber
 &&  \ \ \ \ \  \times \left(\Gamma_{\mu\nu\rho}\Gamma_0\right)_{\alpha\beta}
  \frac{\partial}{\partial\xi_\alpha}\frac{\partial}{\partial\xi_\beta}
  \left(\prod_{\gamma =1}^{16}\xi_\gamma\right),
\end{eqnarray}
where $\zeta_{\mu\nu}(k)$ is a polarization tensor, 
$\zeta_{\mu\nu}(k)=-\zeta_{\nu\mu}(k)$.
Since our SUSY algebra closes only on shell, 
$\tilde{f}_D(k, \xi)$ and $\tilde{f}_B(k, \xi)$ fall into the
supergravity multiplet for the case $k^2=0$. 
Hereafter we, however, formally extend the wave functions to the off-shell 
and integrate over the whole momentum region.  

\subsubsection{Contribution at $\bf {\cal O}(1/d^{8})$ and Myers-like effect}
Let us first look at contributions from the second order terms of $\Phi$.
In these terms we can simply replace $\Phi$ with $\xi$ and thus 
the terms of the order $\tilde{X}^{-3}$ and $\tilde{X}^{-5}$ vanish.
The terms of the order $\tilde{X}^{-7}$,
eq. (\ref{phi^2-x^{-7}}),
 become
\begin{eqnarray}
 \label{xi^2-x^{-7}}
 &&
  \frac{1}{2}
  \left(
  \bar{\xi}\Gamma_{\mu\rho\sigma}\xi
  \right)
  {\cal T}r
  \left[
   -\tilde{F}_{\mu\nu}\frac{1}{\tilde{X}^2}
   \tilde{F}_{\rho\sigma}
   \frac{1}{\tilde{X}^2}
   \tilde{X}_\nu
   \left(\frac{1}{\tilde{X}^2}\right)^2
   -\tilde{F}_{\mu\nu}
   \frac{1}{\tilde{X}^2}
   \tilde{X}_\nu
   \left(\frac{1}{\tilde{X}^2}\right)^2
   \tilde{F}_{\rho\sigma}
   \frac{1}{\tilde{X}^2}
  \right.
  \nonumber \\
 && \hspace{30mm}
  +\tilde{F}_{\mu\nu}
  \left(\frac{1}{\tilde{X}^2}\right)^2
  \tilde{X}_\nu
  \frac{1}{\tilde{X}^2}
  \tilde{F}_{\rho\sigma}
  \frac{1}{\tilde{X}^2}
  +\tilde{F}_{\mu\nu}
  \frac{1}{\tilde{X}^2}
  \tilde{F}_{\rho\sigma}
  \left(\frac{1}{\tilde{X}^2}\right)^2
  \tilde{X}_\nu
  \frac{1}{\tilde{X}^2}
  \nonumber \\
 && \hspace{30mm} \left.
  -\tilde{F}_{\mu\nu}
  \left(\frac{1}{\tilde{X}^2}\right)^2
  \tilde{F}_{\rho\sigma}
  \frac{1}{\tilde{X}^2}
  \tilde{X}_\nu
  \frac{1}{\tilde{X}^2}
  +\tilde{F}_{\mu\nu}
  \frac{1}{\tilde{X}^2}
  \tilde{X}_\nu
  \frac{1}{\tilde{X}^2}
  \tilde{F}_{\rho\sigma}
  \left(\frac{1}{\tilde{X}^2}\right)^2
  \right].
\end{eqnarray} 
We then expand these terms with respect to the inverse powers of
$d_\mu\equiv x_\mu -y_\mu$. 
For example, $1/\tilde{X}^2$ is expanded as follows,
\begin{eqnarray}
 \frac{1}{\tilde{X}^2}=\frac{1}{d^2}\left(1-2\frac{d\cdot A}{d^2}\right)
  + {\cal O}\left(\frac{1}{d^4}\right).
\end{eqnarray}
It is easily realized that the leading terms with $1/d^7$ vanish.
The $1/d^8$ term has the following simple form,
\begin{equation}
 \label{A^5}
 -\frac{1}{2d^8}
  \left(\bar{\xi}\Gamma_{\mu\rho\sigma}\xi\right)
  \mathrm{tr}\left[A_\nu, F_{\mu\nu}\right]F_{\rho\sigma}.
\end{equation}
The $1/d^8$ dependence of the term indicates that the interaction is induced
by an exchange of massless antisymmetric field.  

We then integrate over $y_\mu$ and $\xi$ with the wave function
(\ref{wavefunction-B}) in order to derive the effective action under 
condensation of the antisymmetric tensor field.
In this calculation, we take our wave function (\ref{wavefunction-B}) such
that it damps  at the infrared region where $|y-x| \rightarrow \infty$. 
Such a choice of wave function is natural from the view point of 
the dynamics of the eigenvalues in the matrix model. 
It was indeed shown that the distributions of the eigenvalues
of $A_\mu$ are bounded in a finite region dynamically~\cite{AIKKT}. 
It is therefore natural to consider 
that the wave function damps far from the D-instantons. 
The size of the eigenvalue distribution is a function of $N$.  
If the eigenvalues are distributed 
on $d$-dim hypersurface uniformly, it is proportional to $N^{1/d}$. 
The natural scale of the infrared cutoff of the wave function depends 
on the dynamics of the matrix models, which we do not discuss in the present 
paper. 

The integration over $\xi$ and $y$ can be easily performed as 
\begin{eqnarray}  
&& \int d^{10}y \ d^{16} \xi \ f_B(y, \xi) \ 
 \frac{-1}{2 (x-y)^8} \ \bar{\xi} \Gamma_{\mu \rho \sigma} \xi \ 
 tr [A_\nu, F_{\mu\nu}] F_{\rho\sigma} \nonumber \\ 
&=& \ \int d^{10}y \ d^{10}k \ e^{ik\cdot y} \ 
 \left(
  \zeta_{\mu\nu}(k)k_\rho +\zeta_{\nu\rho}(k)k_\mu +\zeta_{\rho\mu}(k)k_\nu
      \right) 
\ \frac{-1}{2 (x-y)^8} \ tr [A_\sigma, F_{\mu\sigma}] F_{\rho\nu} \nonumber \\
 &=& -\frac{\pi^5}{3} \ \int d^{10}k \  \frac{ e^{ik\cdot x} }{k^2}
  \left(
   \zeta_{\mu\nu}(k)k_\rho +\zeta_{\nu\rho}(k)k_\mu +\zeta_{\rho\mu}(k)k_\nu
   \right) 
\ tr [A_\sigma, F_{\mu\sigma}] F_{\rho\nu}.
\end{eqnarray}
Because of our choice of the wave function, $\zeta_{\mu\nu}(k)$ damps at
small $k$.

We therefore obtain the following effective action 
\begin{equation}
 \label{b-bg}
 S_{eff}(A, x, \psi; f_B) = S_{IKKT} 
  - i\int d^{10}k f_{\mu\nu\rho}(k) e^{ik\cdot x}
  \mathrm{tr}\left[A_\sigma, F_{\mu\sigma}\right]F_{\nu\rho},
\end{equation}
where 
$f_{\mu\nu\rho}(k)=-\frac{i\pi^5}{3}(k_\mu \zeta_{\nu\rho}+k_\nu
\zeta_{\rho\mu} +k_\rho \zeta_{\mu\nu})/k^2$.

This effective action shows that the Chern-Simons-like term is induced by
an effect of condensation of the antisymmetric tensor.
This phenomenon is similar to the Myers effect~\cite{Myers}, but
there is a difference.
In the case of the Myers effect for D0-branes, a cubic term of
bosonic matrices is induced in the RR three-form background. 
This term can be interpreted as a vertex operator for the RR
potential. 
In our case, however, the leading order of the 
induced term in eq.(\ref{b-bg}) is different from 
the expected vertex operator for the charge conjugation of the antisymmetric
tensor field (\ref{Vbc}).
Such a term appears  
at the next order in the $1/d$ expansion as shown in the next subsection.
The reason can be understood as follows. 
If we also calculate the fermionic term containing $\psi$, we would expect
to obtain 
a term like ${\rm tr}\left(\bar{\psi}\Gamma_\mu\psi\right)F_{\nu\rho}$
and the leading order term in  (\ref{b-bg}) with this fermionic term
would be cancelled by using the equation of motion
(\ref{eom-boson}) of the original IKKT action.
This kind of terms can not be seen in the vertex operators 
since we have assumed the equation of motion 
(\ref{eom-boson}) and (\ref{eom-fermion}) in their construction. 
Here,  since we are interested in 
investigating the effective actions under condensation
of the antisymmetric tensor fields,  we do {\it not}  want to use the equations
of motion of the original IKKT action and the term  in (\ref{b-bg}) should
{\it not} be omitted.

Let us see an effect of the induced term in (\ref{b-bg}) for a particular
form of the polarization tensor.
Assuming that the coefficient 
$\int d^{10}k f_{\mu\nu\rho}(k)e^{ik\cdot x}$ is proportional to 
$\epsilon_{ijk}$ with a specific direction $(i,j,k)=(1,2,3)$ and that 
the region $k \sim 0$ is dominant in the $k$-integration, 
the modified matrix model action becomes 
\begin{equation}
 S_{eff}(A, x, \psi; f_B) = 
 S_{IKKT} -  i\alpha \epsilon_{ijk}
  \mathrm{tr}\left[A_\nu, F_{i \nu}\right]F_{jk},
\end{equation}
with a constant coefficient $\alpha$.
This action has a fuzzy sphere classical solution;
\bea
A_i &=& \frac{1}{10\alpha} L_i, \qquad (i = 1, 2, 3) \nonumber \\
A_a &=& 0, \qquad (\mbox{for the other directions}) \nonumber \\
\psi &=& 0.
\eea
The radius of the fuzzy sphere is in inverse proportion
to the coefficient $\alpha$ and in the $\alpha \rightarrow 0$ limit
the fuzzy sphere is expanded and becomes a flat plane.
It contrasts with matrix models with the ordinary cubic Chern-Simons term
(see, for example \cite{Iso-Kimura}) where the radius 
of the fuzzy sphere is proportional to the coefficient of the
Chern-Simons term.

In addition to the fuzzy sphere solution, flat D-branes \begin{equation}
 \left[A_\mu, \ A_\nu\right] = i\theta_{\mu\nu} 1_N
  \qquad \left(\theta_{\mu\nu} = - \theta_{\nu\mu}\right).
\end{equation}
with a constant $\theta_{\mu \nu}$ are also classical solutions 
of the effective action (for an infinite  $N$). It will be
interesting to compare stabilities of these solutions
to the fuzzy sphere solution
by calculating loop corrections around them.

\subsubsection{Contribution at ${\bf {\cal O}(1/d^{9})}$ and $B_{\mu \nu}$ vertex operator}
The induced term in the previous subsection vanishes if we 
use the equation of motion for the configuration $A_{\mu}$.
Then the next order ${\cal O}(1/d^{9})$ 
term becomes the leading order.
From the dimensional analysis, it is expected that the vertex operator
corresponding to the charge conjugation of the antisymmetric tensor field
(\ref{Vbc}) would appear at the order of $1/d^{9}$. 

Expanding the ${\cal O}(\tilde{X}^{-7})$ term (\ref{xi^2-x^{-7}}) with
respect to $1/d$, we obtain  ${\cal O}(1/d^{9})$ terms 
\begin{eqnarray}
 &&
  \frac{2d_\lambda}{d^{10}}
  \left(\bar{\xi}\Gamma_{\mu\rho\sigma}\xi\right)
  {\rm tr}\left(F_{\mu\nu}F_{\rho\sigma}F_{\nu\lambda}
	   +F_{\mu\nu}F_{\nu\lambda}F_{\rho\sigma}
	  \right) \nonumber \\
 &&
 + \frac{2}{d^{10}}
  \left(\bar{\xi}\Gamma_{\mu\rho\sigma}\xi\right)
  {\rm tr}\left(\left[A_\nu, F_{\mu\nu}\right](d\cdot A) F_{\rho\sigma}
	   +\left[A_\nu, F_{\mu\nu}\right]F_{\rho\sigma}(d\cdot A)
	  \right).
\end{eqnarray}
The same order terms with ${\cal O}(1/d^{9})$ can be obtained also from 
eq. (\ref{phi^2-x^{-9}}) as
\begin{eqnarray}
 &&
  -12\frac{d_\lambda}{d^{10}}
  \left(\bar{\xi}\Gamma_{\mu\nu\lambda}\xi\right)
  {\rm tr}\left(F_{\mu\rho}F_{\rho\sigma}F_{\sigma\nu}
	   -\frac{1}{4}F_{\mu\nu}F_{\rho\sigma}F_{\sigma\rho}
	  \right)
  \nonumber \\
 && 
  -2\frac{d_\lambda}{d^{10}}
  \left(\bar{\xi}\Gamma_{\mu\rho\sigma}\xi\right)
  {\rm tr}\left( F_{\mu\nu}F_{\rho\sigma}F_{\nu\lambda}
	   + F_{\mu\nu}F_{\nu\lambda}F_{\rho\sigma}
	  \right).
\end{eqnarray}
Therefore the interaction terms between the mean-field D-instanton and the 
$N\times N$ block are given at this order by
\begin{eqnarray}
  &&
  -12\frac{d_\lambda}{d^{10}}
  \left(\bar{\xi}\Gamma_{\mu\nu\lambda}\xi\right)
  {\rm tr}\left(F_{\mu\rho}F_{\rho\sigma}F_{\sigma\nu}
	   -\frac{1}{4}F_{\mu\nu}F_{\rho\sigma}F_{\sigma\rho}
	  \right)
  \nonumber \\
 && 
  +\frac{2}{d^{10}}
  \left(\bar{\xi}\Gamma_{\mu\rho\sigma}\xi\right)
  {\rm tr}\left(\left[A_\nu, F_{\mu\nu}\right](d\cdot A) F_{\rho\sigma}
	   +\left[A_\nu, F_{\mu\nu}\right]F_{\rho\sigma}(d\cdot A)
	  \right).
\end{eqnarray}
The first line represents an interaction through the vertex operator for
(the charge conjugation of) the antisymmetric tensor field (\ref{Vbc}). 
The second term is similar to the eq.(\ref{A^5}) except for the insertion
of $d\cdot A$.
By integrating over $y_\mu$ and $\xi$ with the wave function
(\ref{wavefunction-B}), the following terms are added to the effective
action, 
\begin{eqnarray}
 &&
  -i\int d^{10}k f_{\mu\nu\rho}(k)e^{ik\cdot x}
  \Bigg\{
  -3i k_\rho \mathrm{tr}
  \left(F_{\mu\sigma}F_{\sigma\lambda}F_{\lambda\nu}
  -\frac{1}{4}F_{\mu\nu}F_{\sigma\lambda}F_{\lambda\sigma}\right)
  \nonumber \\
 && \hspace{35mm}
  +\frac{1}{2}\mathrm{tr}\left(
  \left[A_\sigma, F_{\mu\sigma}\right] \left(ik\cdot A\right)F_{\nu\rho}
  + \left[A_\sigma, F_{\mu\sigma}\right]F_{\nu\rho}\left(ik\cdot A\right) 
  \right)
   \Bigg\}.
\end{eqnarray} 
The first term represents a derivative coupling of D-instantons 
to the vertex operator of  the antisymmetric
tensor field.  
The second term can be combined with eq.(\ref{b-bg}) into a form
\begin{eqnarray}
 - i\int d^{10}k f_{\mu\nu\rho}(k) e^{ik\cdot x}
  \mathrm{Str}~e^{ik\cdot A}
  \left[A_\sigma, F_{\mu\sigma}\right]\cdot F_{\nu\rho}.
\label{expb-bg}
\end{eqnarray}
If we calculate higher order terms in the $1/d$ expansion, 
we would expect to obtain
higher order terms of (\ref{expb-bg}) with respect to
the number of bosonic fields $A_{\mu}$.
\subsection{Condensation of the graviton}
Effects of the condensation of gravitons can be seen from the fourth order
terms of $\xi$.
The term  eq.(\ref{phi^4-x^{-6}}) vanishes by substituting $\xi$ for
$\Phi$ because of the identity for the Majorana-Weyl spinor, 
$\left(\bar{\xi}\Gamma_{\mu\nu\rho}\xi\right)\Gamma^{\nu\rho}\xi=0$. 
Therefore the leading contribution in the $1/d$ expansion comes
from the $\tilde{X}^{-8}$ terms, eq.(\ref{phi^4-x^{-8}}) by replacing 
$\Phi$ with $\xi$ as,
\begin{eqnarray}
 \left(\bar{\xi}\Gamma_{\nu\lambda\rho}\xi\right)
  \left(\bar{\xi}\Gamma_{\mu\rho\sigma}\xi\right)
  {\cal T}r~\frac{1}{\tilde{X}^2}\tilde{F}_{\mu\nu}
  \left(\frac{1}{\tilde{X}^2}\right)^2 \tilde{X}_\lambda
  \left(\frac{1}{\tilde{X}^2}\right)^2 \tilde{X}_\sigma.
\label{xi4-1}
\end{eqnarray}

\subsubsection{Contribution at ${\cal O}(1/d^8)$ and ${\cal O}(1/d^9)$}

Order ${\cal O}(1/d^8)$ terms vanish
\begin{eqnarray}
 \frac{d_\lambda d_\sigma}{d^{10}}
  \left(\bar{\xi}\Gamma_{\nu\lambda\rho}\xi\right)
  \left(\bar{\xi}\Gamma_{\mu\rho\sigma}\xi\right)
  {\rm tr}~F_{\mu\nu}
  = 0,
\end{eqnarray}
since $d_\lambda d_\sigma \left(\bar{\xi}\Gamma_{\nu\lambda\rho}\xi\right)
  \left(\bar{\xi}\Gamma_{\mu\rho\sigma}\xi\right) $
is symmetric under an exchange of $(\mu, \nu)$. 

Similarly order ${\cal O}(1/d^9)$ terms also vanish
\begin{eqnarray}
  \frac{1}{d^{10}}
  \left(\bar{\xi}\Gamma_{\nu\lambda\rho}\xi\right)
  \left(\bar{\xi}\Gamma_{\mu\rho\sigma}\xi\right)
  {\rm tr}\left(d_\lambda F_{\mu\nu}A_\sigma
	   + d_\sigma F_{\mu\nu}A_\lambda\right)
  = 0.
\end{eqnarray}

\subsubsection{Contribution at ${\cal O}(1/d^{10})$}
Hence the leading order terms start from ${\cal O}(1/d^{10})$ terms.
Contributions from the above $\tilde{X}^{-8}$ term (\ref{xi4-1}) are given by 
\begin{eqnarray}
 \label{graviton-1}
 \frac{1}{2d^{10}}
  \left(\bar{\xi}\Gamma_{\mu\rho\lambda}\xi\right)
  \left(\bar{\xi}\Gamma_{\nu\sigma\lambda}\xi\right)
  {\rm tr}~F_{\mu\nu}F_{\rho\sigma}
  +\frac{4}{d^{12}}d_\lambda
  \left(\bar{\xi}\Gamma_{\mu\rho\sigma}\xi\right)
  c_{\mu\nu}(\xi){\rm tr}~F_{\nu\rho}F_{\sigma\lambda},
\end{eqnarray}
where 
$c_{\mu\nu}(\xi)\equiv d_\rho\left(\bar{\xi}\Gamma_{\mu\nu\rho}\xi\right)$.
The same order terms are also obtained from the 
$\tilde{X}^{-10}$ terms (\ref{phi^4-x^{-10}}) as
\begin{eqnarray}
 \label{graviton-2}
 &&
  -\frac{1}{8d^{10}}
  \left(\bar{\xi}\Gamma_{\mu\nu\lambda}\xi\right)
  \left(\bar{\xi}\Gamma_{\rho\sigma\lambda}\xi\right)
  {\rm tr}~F_{\mu\nu}F_{\rho\sigma}
  -\frac{1}{2d^{12}}c_{\mu\rho}c_{\rho\nu}
  {\rm tr}~F_{\mu\sigma}F_{\sigma\nu}
  +\frac{3}{2d^{12}}c_{\mu\nu}c_{\rho\sigma}
  {\rm tr}~F_{\mu\nu}F_{\rho\sigma}
  \nonumber \\
 &&
  -\frac{9}{2d^{12}}c_{\mu\rho}c_{\nu\sigma}
  {\rm tr}~F_{\mu\nu}F_{\rho\sigma}
  +\frac{3}{2d^{12}}d_\lambda c_{\rho\sigma}
  \left(\bar{\xi}\Gamma_{\mu\nu\rho}\xi\right)
  {\rm tr}~F_{\mu\nu}F_{\sigma\lambda}.
\end{eqnarray}
By using the following Fierz identity,
\begin{eqnarray}
 c_{\mu\nu}c_{\rho\sigma} &=&
  \frac{1}{3}\left(c_{\mu\nu}c_{\rho\sigma}+c_{\mu\rho}c_{\sigma\nu}
	      +c_{\mu\sigma}c_{\nu\rho}\right)
  \nonumber \\
 && -\frac{1}{6}
  \left(g_{\mu\rho}c_{\nu\lambda}c_{\lambda\sigma}
   -g_{\mu\sigma}c_{\nu\lambda}c_{\lambda\rho}
   -g_{\nu\rho}c_{\mu\lambda}c_{\lambda\sigma}
   +g_{\nu\sigma}c_{\mu\lambda}c_{\lambda\rho}
  \right)
  \nonumber \\
 && +\frac{1}{6}
  \left[
   d_\mu c_{\lambda\nu}\left(\bar{\xi}\Gamma_{\rho\sigma\lambda}\xi\right)
   -d_\nu c_{\lambda\mu}\left(\bar{\xi}\Gamma_{\rho\sigma\lambda}\xi\right)
   +d_\rho c_{\lambda\sigma}\left(\bar{\xi}\Gamma_{\mu\nu\lambda}\xi\right)
   -d_\sigma c_{\lambda\rho}\left(\bar{\xi}\Gamma_{\mu\nu\lambda}\xi\right)
  \right]
  \nonumber \\
 && +\frac{d^2}{6}\left(\bar{\xi}\Gamma_{\mu\nu\lambda}\xi\right)
  \left(\bar{\xi}\Gamma_{\rho\sigma\lambda}\xi\right).
\end{eqnarray}
the sum of these two terms, eqs.(\ref{graviton-1}) and
(\ref{graviton-2}), can be simplified and depends on $\xi$
only in the form of $c_{\mu\nu}(\xi)$ as
\begin{eqnarray}
 (\ref{graviton-1})+(\ref{graviton-2})
  = -\frac{1}{d^{12}}c_{\mu\rho}c_{\rho\nu}
  {\rm tr}~F_{\mu\sigma}F_{\sigma\nu}.
\end{eqnarray}
It represents a derivative coupling of a single D-instanton 
to the graviton vertex operator constructed from 
the $N$ D-instantons.
If we insert the graviton wave function and integrate over the single
D-instanton coordinates, we can obtain the graviton vertex operator
as an induced term in the effective action.

Similarly  interactions  mediated by the 4-th rank self-dual
antisymmetric tensor field would appear, but such terms vanish in the
leading order because of the cyclic property of  
the trace and the Jacobi identity,
\begin{eqnarray}
 && 
  c_{\mu\nu}c_{\rho\sigma}
  {\rm tr}\left(F_{\mu\nu}F_{\rho\sigma}
	   + F_{\mu\rho}F_{\sigma\nu} + F_{\mu\sigma}F_{\nu\rho}
	  \right) 
  \nonumber \\
 && \hspace{10mm}
  = c_{\mu\nu}c_{\rho\sigma}
  {\rm tr}A_\mu
  \left(\left[A_\nu, \left[A_\rho, A_\sigma\right]\right]
   + \left[A_\rho, \left[A_\sigma, A_\nu\right]\right]
   + \left[A_\sigma, \left[A_\nu, A_\rho\right]\right]
  \right)
  = 0.
\end{eqnarray}
If we calculate higher order terms, we would expect to obtain 
the terms which can be produced by expanding the exponential in   
\be
  \frac{1}{d^{12}}c_{\mu\nu}c_{\rho\sigma} \
  {\rm Str} \ e^{i k \cdot A} \left(F_{\mu\nu}\cdot F_{\rho\sigma}
	   + F_{\mu\rho}\cdot F_{\sigma\nu} 
	   + F_{\mu\sigma}\cdot F_{\nu\rho}
	  \right) .
 \ee

\section{Conclusion}
\setcounter{equation}{0}
In this paper, we have considered fermionic backgrounds and condensation of
supergravity fields in the IIB matrix model.
We start from the type IIB matrix model in a flat background
with the size $(N+1)\times(N+1)$, namely a system of
$(N+1)$ D-instantons.
We then integrate one D-instanton (which we call a mean-field D-instanton)
and obtain an effective action for $N$ D-instantons by assuming particular
forms of wave functions of the mean-field D-instanton.
If we assume that the configurations of $N$ D-instantons satisfy
the equation of motion, we show that vertex operators obtained 
in our previous paper \cite{ITU} are induced in the effective action 
as leading contributions.
If we do not assume it, extra terms also appear.
In particular if we take the wave function as that of the antisymmetric
tensor field, a Chern-Simons like term is induced in the leading order of
perturbations.
Though this term is quintic with respect to the field $A_{\mu}$, 
a fuzzy sphere becomes a solution to the equation of motion.
In this sense, this is a similar mechanism to the Myers effect.

We have also given a stringy interpretation of the wave functions
of the mean-field D-instanton as overlaps of the D-instanton boundary
state with closed string massless states.
The ordinary D-instanton only couples with the dilaton and the axion 
states. But since a D-instanton is a half-BPS state and breaks 
one half of the supersymmetries, we can obtain other states
by acting broken supersymmetry generators on the ordinary D-instanton
state. They couple to other supergravity fields through derivative couplings
and form a supersymmetry multiplet in type IIB supergravity.  
We showed that the wave functions are nothing but the overlaps
of these D-instanton boundary states with massless closed string states.

It is interesting to investigate effective actions under condensation of
every massless closed string mode systematically, besides the charge
conjugation of the antisymmetric tensor field and graviton we studied in
this paper. Though it is expected from the analysis of the string theory
side that each mode couples to the vertex operator of the $N$ D-instanton
system through an appropriate derivative coupling, other types of couplings
like the quintic term derived here can also appear.
We think that such studies clarify how the IIB matrix model contains dynamics
of closed strings.

\setcounter{equation}{0}

\section*{Acknowledgements}
We would like to thank H. Aoki, K. Hamada, Y. Kimura, Y. Kitazawa, 
J. Nishimura, T. Suyama, D. Tomino, A. Tsuchiya and K. Yoshida 
for useful discussions. 
We would especially thank K. Yoshida for discussions on section 3.

\section*{Appendix}
\renewcommand{\theequation}{A.\arabic{equation}}
\setcounter{equation}{0}
In this appendix, we briefly review the boundary states in the Green-Schwarz 
formalism of type IIB superstrings in the light-cone gauge.

We first summarize our notations: 
\begin{itemize}
 \item \underline{Space-time quantities (in $(9+1)$-dimensions)}  \\
       Metric: 
       \begin{equation}
	\eta_{\mu\nu}=diag(-1, +1, \cdots, +1),  	
       \end{equation}
       Gamma matrices (in Majorana representation): 
       \begin{eqnarray}
	&& \{\Gamma^\mu, \Gamma^\nu\}=-2\eta^{\mu\nu}, \\
	&& \Gamma^0 = \sigma_2\otimes 1_{16}, \\
	&& \Gamma^i = i\sigma_1\otimes \gamma^i, \ \ (i=1,2, \cdots, 8) \\
	&& \Gamma^9 = i\sigma_3\otimes 1_{16}, \\
	&& \Gamma_{11} = \Gamma^0\Gamma^1\cdots\Gamma^9
	 = -\sigma_1\otimes 
	 \left(
	  \begin{array}{cc}
	   1_{8} & 0 \\
	   0 & -1_{8}
	  \end{array}
	      \right), \\
	&&\gamma^i = 
	 \left(
	  \begin{array}{cc}
	   0 & \gamma^i_{a\dot{a}}\\
	   \gamma^i_{\dot{a}a} & 0
	  \end{array}
	    \right), 
	 \qquad \gamma^i_{a\dot{a}}=\gamma^i_{\dot{a}a}, \\
	&& \gamma^i_{a\dot{a}}\gamma^j_{\dot{a}b}
	 +\gamma^j_{a\dot{a}}\gamma^i_{\dot{a}b}
	 = 2\delta^{ij}\delta_{ab}, \\
	&& \gamma^i_{a\dot{a}}\gamma^i_{b\dot{b}}
	 +\gamma^i_{b\dot{a}}\gamma^i_{a\dot{b}}
	 = 2\delta_{ab}\delta_{\dot{a}\dot{b}}, 
       \end{eqnarray}
       Spinors: 
       \begin{eqnarray}
	&& \theta =
	 \left(
	  \theta_1^a, \theta_1^{\dot{a}},
	  \theta_2^a, \theta_2^{\dot{a}} \right)^T,
       \end{eqnarray}
       Weyl spinors:
       \begin{eqnarray}
	&& \Gamma_{11} \theta = \theta
	 \ \Longrightarrow \ 
	 \theta = 
	 \left(\theta^a, \theta^{\dot{a}},
	  -\theta^a, \theta^{\dot{a}} \right)^T, \\
	&& \Gamma_{11} \theta = -\theta
	 \ \Longrightarrow \ 
	 \theta = 
	 \left(\theta^a, \theta^{\dot{a}},
	  \theta^a, -\theta^{\dot{a}} \right)^T,
       \end{eqnarray}
 \item \underline{World-sheet quantities}\\
       Metric:
       \begin{equation}
	\eta_{\alpha\beta}=diag(-1, +1)	
       \end{equation}
       Gamma matrices:
       \begin{eqnarray}
	&& \{\rho^\alpha, \rho^\beta\}= -2\eta^{\alpha\beta}, \\
	&& \rho^0 = \sigma_2, \quad 
	 \rho^1 = i\sigma_1,
       \end{eqnarray}
       Antisymmetric tensor $\epsilon^{\alpha\beta}$:
       \begin{equation}
	\epsilon^{01}=+1,
       \end{equation}
\end{itemize}

In the Green-Schwarz formalism, the IIB superstring theory is described
by ten real bosons $X^\mu ~(\mu=0, 1, \cdots, 9)$ and two Majorana-Weyl
fermions $\theta^A ~(A=1, 2)$ with the same chirality
$\Gamma_{11}\theta^A=-\theta^A$. 
Here we take the light-cone gauge;
\begin{eqnarray}
 \Gamma^+ \theta^A &=& 0 \quad  \longrightarrow \quad 
  \theta^A = \left(\theta^{Aa}, 0, \theta^{Aa}, 0\right), \\
 X^+ &=& x^+ + p^+ \tau.
\end{eqnarray}
The light-cone components are defined as
\begin{eqnarray}
 \Gamma^{\pm} &=& \frac{1}{\sqrt{2}}\left(\Gamma^0 \pm \Gamma^9\right), \\
 X^{\pm} &=& \frac{1}{\sqrt{2}}\left(X^0 \pm X^9\right).
\end{eqnarray}
The explicit forms of $\Gamma^{\pm}$ are
\begin{eqnarray}
 \Gamma^+ &=& \frac{i}{\sqrt{2}}
  \left(
   \begin{array}{cc}
    1 & -1 \\
    1 & -1 
   \end{array}
  \right)
  \otimes 1_{16}, \\
 \Gamma^- &=& \frac{i}{\sqrt{2}}
  \left(
   \begin{array}{cc}
    -1 & -1 \\
    1 & 1 
   \end{array}
  \right)
  \otimes 1_{16}.
\end{eqnarray}

The world-sheet action in the light-cone gauge is given by
\begin{eqnarray}
 \label{lc-action}
 S_{l.c.} &=& -\frac{1}{4\pi}\int d^2\sigma
  \left(
   \partial_\alpha X^i \partial^\alpha X^i 
   -i \bar{S}^a \rho^\alpha \partial_\alpha S^a
  \right) \nonumber \\
 &=& -\frac{1}{4\pi}\int d^2\sigma
  \left[
   -\left(\partial_\tau X^i\right)^2 + \left(\partial_\sigma X^i\right)^2
   -iS^{1a}\left(\partial_\tau + \partial_\sigma\right)S^{1a}
   -iS^{2a}\left(\partial_\tau - \partial_\sigma\right)S^{2a}
  \right], \nonumber \\
\end{eqnarray}
where $S^{Aa}$ are proportional to $\theta^{Aa}$;
$S^{Aa} \propto \sqrt{p^+}\theta^{Aa}$.
The coordinates are expanded with respect to the Fourier modes as 
\begin{eqnarray}
 X^i &=& x^i + p^i \tau 
  +\frac{i}{\sqrt{2}}\sum_{n\neq 0}\frac{1}{n}
  \left(
   \alpha^i_n e^{-in(\tau-\sigma)}+\tilde{\alpha}^i_n e^{-in(\tau+\sigma)}
  \right), \\
 S^{1a} &=& \sum_{n}S^a_n e^{-in(\tau-\sigma)}, \\
 S^{2a} &=& \sum_{n}\tilde{S}^a_n e^{-in(\tau+\sigma)}.
\end{eqnarray}
Under the quantization, the mode operators satisfy the hermiticity
conditions
\begin{equation}
 \alpha^\dagger_n = \alpha_{-n}, \quad
  \tilde{\alpha}^\dagger_n = \tilde{\alpha}_{-n}, \quad
  \left(S^a_n\right)^\dagger = S^a_{-n}, \quad
  \left(\tilde{S}^a_n\right)^\dagger = \tilde{S}^a_{-n}.
\end{equation}
Also the commutation relations among them are given by
\begin{eqnarray}
 && [x^i, p^j] = i\delta^{ij}, \qquad
  [\alpha^i_m, \alpha^j_n] = m\delta^{ij}\delta_{m+n,0}, \qquad
  [\tilde{\alpha}^i_m, \tilde{\alpha}^j_n] = m\delta^{ij}\delta_{m+n,0}, \\
 && \{S^a_m, S^b_n\} = \delta^{ab}\delta_{m+n,0}, \qquad
  \{\tilde{S}^a_m, \tilde{S}^b_n\} = \delta^{ab}\delta_{m+n,0}.
\end{eqnarray}

The action (\ref{lc-action}) has the ${\cal N}=2$ supersymmetry
consisting of the kinematical SUSY
\begin{eqnarray}
 \delta S^{Aa} &=& \sqrt{2p^+} \epsilon^{Aa}, \\
 \delta X^i &=& 0,
\end{eqnarray}
and the dynamical SUSY
\begin{eqnarray}
 \delta S^{1a} &=& \frac{1}{\sqrt{p^+}}
  \left(\partial_\tau - \partial_\sigma\right)X^i 
  \gamma^i_{a\dot{a}}\epsilon^{1\dot{a}}, \\
\delta S^{2a} &=& \frac{1}{\sqrt{p^+}}
  \left(\partial_\tau + \partial_\sigma\right)X^i 
  \gamma^i_{a\dot{a}}\epsilon^{2\dot{a}}, \\
 \delta X^i &=& -\frac{i}{\sqrt{p^+}}\epsilon^{A\dot{a}}
  \gamma^i_{\dot{a}a}S^{Aa}.
\end{eqnarray}
These transformations are generated by the following supercharges
\begin{eqnarray}
 Q^{1a} &=& \int_0^{2\pi}\frac{d\sigma}{2\pi}\sqrt{2p^+}S^{1a} 
  = \sqrt{2p^+}S^{a}_0, \\
 Q^{2a} &=& \int_0^{2\pi}\frac{d\sigma}{2\pi}\sqrt{2p^+}S^{2a} 
  = \sqrt{2p^+}\tilde{S}^{a}_0, \\
 Q^{1\dot{a}} &=& \int_0^{2\pi}\frac{d\sigma}{2\pi}
  \frac{1}{\sqrt{p^+}}\left(\partial_\tau - \partial_\sigma\right)X^i
  \gamma^i_{\dot{a}a}S^{1a}
  = \frac{1}{\sqrt{p^+}}\gamma^i_{\dot{a}a}
  \left(p^i S^a_0 + \sqrt{2}\sum_{n\neq 0}\alpha^i_n S^a_{-n}\right), \\
 Q^{2\dot{a}} &=& \int_0^{2\pi}\frac{d\sigma}{2\pi}
  \frac{1}{\sqrt{p^+}}\left(\partial_\tau + \partial_\sigma\right)X^i
  \gamma^i_{\dot{a}a}S^{2a}
  = \frac{1}{\sqrt{p^+}}\gamma^i_{\dot{a}a}
  \left(p^i \tilde{S}^a_0 
   + \sqrt{2}\sum_{n\neq 0}\tilde{\alpha}^i_n \tilde{S}^a_{-n}\right),
\end{eqnarray}
which satisfy the algebra
\begin{eqnarray}
 \{Q^{Aa}, Q^{Bb}\} &=& 2p^+ \delta^{AB}\delta^{ab}, \\
 \{Q^{1\dot{a}}, Q^{1\dot{b}}\} &=& 2P^- \delta_{\dot{a}\dot{b}}, \\
 \{Q^{2\dot{a}}, Q^{2\dot{b}}\} &=& 2\tilde{P}^- \delta_{\dot{a}\dot{b}}, \\
 \{Q^{Aa}, Q^{B\dot{a}}\} &=& \sqrt{2}\gamma^i_{a\dot{a}}p^i \delta^{AB},
\end{eqnarray}
with
\begin{eqnarray}
 P^- &=& \frac{1}{p^+}
  \left[
   \frac{p^i p^i}{2}+\sum_{n\neq 0} 
   \left(nS^a_{-n}S^a_n + \alpha^i_{-n}\alpha^i_n\right)
  \right], \\
 \tilde{P}^- &=& \frac{1}{p^+}
  \left[
   \frac{p^i p^i}{2}+\sum_{n\neq 0} 
   \left(n\tilde{S}^a_{-n}\tilde{S}^a_n 
    + \tilde{\alpha}^i_{-n}\tilde{\alpha}^i_n\right)
  \right].
\end{eqnarray}

A boundary state is usually defined by a set of the boundary conditions on a
constant $\tau$ surface:
\begin{eqnarray}
 && \left[(\partial_\tau - \partial_\sigma)X^i
     -M_{ij}(\partial_\tau + \partial_\sigma)X^j\right]
 |B, \eta \rangle = 0, 
 \label{bc-1A}\\
 && Q^{+a}_\eta |B, \eta \rangle \equiv 
  \left(Q^{1a}+i\eta M_{ab} Q^{2b}\right)|B, \eta \rangle = 0, 
  \label{bc-2A}\\
 && Q^{+\dot{a}}_\eta |B, \eta \rangle \equiv 
  \left(Q^{1\dot{a}}+i\eta M_{\dot{a}\dot{b}} Q^{2\dot{b}}\right)
  |B, \eta \rangle = 0,
  \label{bc-3A}
\end{eqnarray}
where $\eta$ is a parameter $(\eta^2 = 1)$, and $M_{ij}$ is an element of
$SO(8)$. 
For the Neumann directions $M_{ij}=-\delta_{ij}$ and for the Dirichlet
directions $M_{ij}=\delta_{ij}$. 
$M_{ab}$ and $M_{\dot{a}\dot{b}}$ are determined by consistency requirements
as follows.
Taking the surface $\tau =0$, the conditions (\ref{bc-1A})-(\ref{bc-3A}) 
are written in terms of the mode operators as 
\begin{eqnarray}
 && \left(p^i - M_{ij} p^j\right) |B, \eta \rangle = 0,
  \label{bc-10A} \\
 && \left(\alpha^i_n - M_{ij} \tilde{\alpha}^j_{-n}\right)
 |B, \eta \rangle = 0, 
  \label{bc-1nA} \\
 && \left(S^a_0 + i\eta M_{ab}\tilde{S}^b_0\right)
  |B, \eta \rangle = 0, 
  \label{bc-2nA}\\
 && \left[\gamma^i_{\dot{a}a}p^i S^a_0 
     + i\eta M_{\dot{a}\dot{b}}\gamma^i_{\dot{b}a}\tilde{S}^a_0
    +\sqrt{2}\sum_{n\neq 0}
    \left(\gamma^i_{\dot{a}a}\alpha^i_n S^a_{-n}
    + i\eta M_{\dot{a}\dot{b}}\gamma^i_{\dot{b}a}
    \tilde{\alpha}^i_n\tilde{S}^a_{-n}\right)
    \right]
 |B, \eta \rangle = 0.
 \label{bc-3nA} \nonumber \\
\end{eqnarray}
Let us determine $M_{ab}$ and $M_{\dot{a}\dot{b}}$. 
From $\{Q^{+a}, Q^{+b}\}|B, \eta \rangle = 0$, we find
\begin{equation}
 M_{ac} M_{bc} = \delta_{ab}, 
\end{equation}
meaning that $M_{ab}$ is an orthogonal matrix.
Next, $\{Q^{+a}, Q^{+\dot{a}}\}|B, \eta \rangle = 0$ leads to
\begin{equation}
 \left(\gamma^i_{a\dot{a}}p^i
 -M_{ab}M_{\dot{a}\dot{b}}\gamma^i_{\dot{b}b}p^i\right)
 |B, \eta \rangle = 0.
\end{equation} 
Comparing this with (\ref{bc-10A}), we have
\begin{equation}
 \gamma^i_{a\dot{a}} M_{ij} 
  - M_{ab}M_{\dot{a}\dot{b}}\gamma^j_{\dot{b}b} = 0.
  \label{relation-M}
\end{equation}
The consistency between eq.(\ref{bc-1A}) and eq.(\ref{bc-3A}) requires 
\begin{equation}
 \left(\gamma^i_{\dot{a}a}S^a_n
 + i\eta M_{ij}M_{\dot{a}\dot{b}}\gamma^j_{\dot{b}b}\tilde{S}^b_{-n}\right)
 |B, \eta \rangle = 0  \qquad \mbox{for } n\neq 0,
\end{equation} 
by using (\ref{relation-M}), which are rewritten as
\begin{equation}
 \gamma^i_{\dot{a}a}\left(S^a_n + i\eta M_{ab}\tilde{S}^b_{-n}\right)
  |B, \eta \rangle = 0 \qquad \mbox{for } n\neq 0.
\end{equation}  

Since $M_{ij}$ is an element of $SO(8)$, it can be written as 
$M_{ij}=\left(e^{\Omega_{kl}\Sigma^{kl}}\right)_{ij}$ with 
$\left(\Sigma^{kl}\right)_{ij}=\delta^k_i\delta^l_j-\delta^l_i\delta^k_j$
being generators of $SO(8)$.
Eq.(\ref{relation-M}) can be solved in terms of $\Omega_{ij}$ as
\begin{eqnarray}
 M_{ab} &=& \left(e^{\frac{1}{2}\Omega_{ij}\gamma^{ij}}\right)_{ab}, \\
 M_{\dot{a}\dot{b}} &=& 
  \left(e^{\frac{1}{2}\Omega_{ij}\tilde{\gamma}^{ij}}\right)_{\dot{a}\dot{b}},
\end{eqnarray}
where
\begin{equation}
 \gamma^{ij}_{ab} = \frac{1}{2}
  \left(
   \gamma^i_{a\dot{a}}\gamma^j_{\dot{a}b}
   -\gamma^j_{a\dot{a}}\gamma^i_{\dot{a}b}
  \right), 
  \qquad
  \tilde{\gamma}^{ij}_{\dot{a}\dot{b}} = \frac{1}{2}
  \left(
   \gamma^i_{\dot{a}a}\gamma^j_{a\dot{b}}
   -\gamma^j_{\dot{a}a}\gamma^i_{a\dot{b}}
  \right).
\end{equation}

The boundary state $|B, \eta\rangle$ can be expressed in the form
\begin{equation}
 \label{boundary-state}
 |B, \eta \rangle =
  e^{\sum_{n>0}
  \left(
   \frac{1}{n}M_{ij}\alpha^i_{-n}\tilde{\alpha}^j_{-n}
   -i\eta M_{ab}S^a_{-n}\tilde{S}^b_{-n}
  \right)}
  |B_0, \eta \rangle
\end{equation} 
with the zero-mode part
\begin{equation}
 \label{bs-zero}
 |B_0, \eta \rangle 
  = C\left(M_{ij} |i\rangle |j\rangle
     -i\eta M_{\dot{a}\dot{b}}|\dot{a}\rangle |\dot{b}\rangle\right).
\end{equation}
$C$ is a normalization constant, and the ground states $|i\rangle$ and
$|\dot{a}\rangle$ are defined by 
\begin{eqnarray}
 && \alpha^j_n |i\rangle = S^a_n |i\rangle 
  = \alpha^i_n |\dot{a}\rangle = S^a_n |\dot{a}\rangle = 0, 
  \qquad (\mbox{for} \ n>0), \\
 && S^a_0 |i\rangle =
  \frac{\gamma^i_{a\dot{a}}}{\sqrt{2}}|\dot{a}\rangle, \qquad 
  S^a_0 |\dot{a}\rangle = \frac{\gamma^i_{a\dot{a}}}{\sqrt{2}}|i\rangle.
\end{eqnarray}

Broken supercharges are given by
\begin{eqnarray}
 \label{broken-charge1}
 Q^{-a}_\eta &\equiv& Q^{1a}-i\eta M_{ab} Q^{2b}, \\
 Q^{-\dot{a}}_\eta &\equiv& 
  Q^{1\dot{a}}-i\eta M_{\dot{a}\dot{b}} Q^{2\dot{b}},
  \label{broken-charge2}
\end{eqnarray}
and the algebra of broken and unbroken supercharges becomes
\begin{eqnarray}
 \{Q^{+a}_\eta, Q^{-b}_\eta\} &=& 4p^+ \delta_{ab}, \\
 \{Q^{+a}_\eta, Q^{-\dot{b}}_\eta\} &=& 
  \sqrt{2}\gamma^i_{a\dot{b}}\left(p^i+M_{ij}p^j\right), \\
 \{Q^{+\dot{a}}_\eta, Q^{-b}_\eta\} &=&
  \sqrt{2}\gamma^i_{\dot{a}b}\left(p^i+M_{ij}p^j\right), \\
 \{Q^{+\dot{a}}_\eta, Q^{-\dot{b}}_\eta\} &=&
  2(P^- + \tilde{P}^-) \delta_{\dot{a}\dot{b}}
  =2P^-_{cl} \delta_{\dot{a}\dot{b}},
\end{eqnarray}
and the other anticommutators vanish.

\end{document}